\documentclass[trackchanges, twocolumn, twocolappendix]{aastex701}

\usepackage{enumitem}

\usepackage{color}

\shorttitle{Voids from Galaxy Group Samples}
\shortauthors{Y.Song et al.}
\graphicspath{{./}{figures/}}
\usepackage{subfigure} 
\usepackage{amsmath}
\begin{document}

\title{CSST large-scale structure analysis pipeline: IV. Cosmic Voids Identified from Galaxy Group Samples as Probes of the Large-scale Structure}
\author[orcid=0009-0007-5626-9489]{Yingxiao Song}
\affiliation{Department of Astronomy, School of Physics and Astronomy, Shanghai Jiao Tong University, Shanghai 200240, People’s Republic of China}
\affiliation{Shanghai Key Laboratory for Particle Physics and Cosmology, and Key Laboratory for Particle Physics, Astrophysics and Cosmology, Ministry of Education, Shanghai Jiao Tong University, Shanghai 200240, People’s Republic of China}
\email[show]{yxsong@sjtu.edu.cn}  

\author[orcid=0000-0003-3997-4606]{Xiaohu Yang}
\affiliation{Department of Astronomy, School of Physics and Astronomy, Shanghai Jiao Tong University, Shanghai 200240, People’s Republic of China}
\affiliation{Shanghai Key Laboratory for Particle Physics and Cosmology, and Key Laboratory for Particle Physics, Astrophysics and Cosmology, Ministry of Education, Shanghai Jiao Tong University, Shanghai 200240, People’s Republic of China}
\affiliation{State Key Laboratory of Dark Matter Physics, Tsung-Dao Lee Institute, Shanghai Jiao Tong University, Shanghai 200240, People’s Republic of China}
\email[show]{xyang@sjtu.edu.cn}  

\author[orcid=0000-0003-0709-0101]{Yan Gong}
\affiliation{National Astronomical Observatories, Chinese Academy of Sciences,20A Datun Road, Beijing 100012, People’s Republic of China}
\affiliation{School of Astronomy and Space Sciences, University of Chinese Academy of Sciences(UCAS),\\Yuquan Road NO.19A Beijing 100049, People’s Republic of China}
\affiliation{Science Center for China Space Station Telescope, National Astronomical Observatories, Chinese Academy of Sciences,\\20A Datun Road, Beijing 100101, People’s Republic of China}
\email{gongyan@bao.ac.cn}

\author[orcid=0000-0003-3196-7938]{Yizhou Gu}
\affiliation{Department of Astronomy, School of Physics and Astronomy, Shanghai Jiao Tong University, Shanghai 200240, People’s Republic of China}
\affiliation{Shanghai Key Laboratory for Particle Physics and Cosmology, and Key Laboratory for Particle Physics, Astrophysics and Cosmology, Ministry of Education, Shanghai Jiao Tong University, Shanghai 200240, People’s Republic of China}
\affiliation{State Key Laboratory of Dark Matter Physics, Tsung-Dao Lee Institute, Shanghai Jiao Tong University, Shanghai 200240, People’s Republic of China}
\email{guyizhou@sjtu.edu.cn}

\author[orcid=0000-0003-0771-1350]{Qingyang Li}
\affiliation{Department of Astronomy, School of Physics and Astronomy, Shanghai Jiao Tong University, Shanghai 200240, People’s Republic of China}
\affiliation{Shanghai Key Laboratory for Particle Physics and Cosmology, and Key Laboratory for Particle Physics, Astrophysics and Cosmology, Ministry of Education, Shanghai Jiao Tong University, Shanghai 200240, People’s Republic of China}
\email{qingyli@sjtu.edu.cn}

\author[orcid=0000-0003-4936-8247]{Hong Guo}
\affiliation{Shanghai Astronomical Observatory, Chinese Academy of Sciences, 80 Nandan Road, Shanghai 200030, People’s Republic of China}
\email{guohong@shao.ac.cn}

\author[orcid=0000-0002-2547-0434]{Yunkun Han}
\affiliation{International Centre of Supernovae (ICESUN), Yunnan Key Laboratory of Supernova Research, Yunnan Observatories, Chinese Academy of Sciences, Kunming 650216, People’s Republic of China}
\affiliation{Center for Astronomical Mega-Science, Chinese Academy of Sciences, 20A Datun Road, Chaoyang District, Beijing 100012, People’s Republic of China}
\email{hanyk@ynao.ac.cn}

\author[orcid=0000-0002-4534-3125]{Yipeng Jing}
\affiliation{Department of Astronomy, School of Physics and Astronomy, Shanghai Jiao Tong University, Shanghai 200240, People’s Republic of China}
\affiliation{Shanghai Key Laboratory for Particle Physics and Cosmology, and Key Laboratory for Particle Physics, Astrophysics and Cosmology, Ministry of Education, Shanghai Jiao Tong University, Shanghai 200240, People’s Republic of China}
\affiliation{State Key Laboratory of Dark Matter Physics, Tsung-Dao Lee Institute, Shanghai Jiao Tong University, Shanghai 200240, People’s Republic of China}
\email{ypjing@sjtu.edu.cn}  

\author[orcid=0000-0002-8711-8970]{Cheng Li}
\affiliation{Department of Astronomy, Tsinghua University, Beijing 100084, People’s Republic of China}
\email{cli2015@tsinghua.edu.cn}

\author[orcid=0000-0002-9968-2894]{Feng Shi}
\affiliation{School of Aerospace Science and Technology, Xidian University, Xi’an 710126, People’s Republic of China}
\email{fshi@xidian.edu.cn}

\author[orcid=0000-0003-2132-0727]{Jipeng Sui}
\affiliation{School of Astronomy and Space Sciences, University of Chinese Academy of Sciences(UCAS),\\Yuquan Road NO.19A Beijing 100049, People’s Republic of China}
\affiliation{Key Laboratory of Optical Astronomy, National Astronomical Observatories, Chinese Academy of Sciences, Beijing 100101, People’s Republic of China}
\email{suijp@bao.ac.cn}

\author[orcid=0000-0002-8705-6327]{Run Wen}
\affiliation{Department of Astronomy, School of Physics and Astronomy, Shanghai Jiao Tong University, Shanghai 200240, People’s Republic of China}
\affiliation{Shanghai Key Laboratory for Particle Physics and Cosmology, and Key Laboratory for Particle Physics, Astrophysics and Cosmology, Ministry of Education, Shanghai Jiao Tong University, Shanghai 200240, People’s Republic of China}
\affiliation{State Key Laboratory of Dark Matter Physics, Tsung-Dao Lee Institute, Shanghai Jiao Tong University, Shanghai 200240, People’s Republic of China}
\email{wenrun1214@sjtu.edu.cn}

\author[orcid=0000-0003-1718-6481]{Hu Zhan}
\affiliation{School of Astronomy and Space Sciences, University of Chinese Academy of Sciences(UCAS),\\Yuquan Road NO.19A Beijing 100049, People’s Republic of China}
\affiliation{Key Laboratory of Optical Astronomy, National Astronomical Observatories, Chinese Academy of Sciences, Beijing 100101, People’s Republic of China}
\email{zhanhu@nao.cas.cn}

\author[orcid=0000-0003-2632-9915]{Pengjie Zhang}
\affiliation{Department of Astronomy, School of Physics and Astronomy, Shanghai Jiao Tong University, Shanghai 200240, People’s Republic of China}
\affiliation{Shanghai Key Laboratory for Particle Physics and Cosmology, and Key Laboratory for Particle Physics, Astrophysics and Cosmology, Ministry of Education, Shanghai Jiao Tong University, Shanghai 200240, People’s Republic of China}
\affiliation{State Key Laboratory of Dark Matter Physics, Tsung-Dao Lee Institute, Shanghai Jiao Tong University, Shanghai 200240, People’s Republic of China}
\email{zhangpj@sjtu.edu.cn}

\author[orcid=0000-0003-1967-4091]{Youcai Zhang}
\affiliation{Shanghai Astronomical Observatory, Chinese Academy of Sciences, 80 Nandan Road, Shanghai 200030, People’s Republic of China}
\email{yczhang@shao.ac.cn}

\author[orcid=0000-0003-4726-6714]{Gong-Bo Zhao}
\affiliation{National Astronomical Observatories, Chinese Academy of Sciences,20A Datun Road, Beijing 100012, People’s Republic of China}
\affiliation{School of Astronomy and Space Sciences, University of Chinese Academy of Sciences(UCAS),\\Yuquan Road NO.19A Beijing 100049, People’s Republic of China}
\email{gbzhao@nao.cas.cn}

\author[orcid=0000-0003-3728-9912]{Xian Zhong Zheng}
\affiliation{State Key Laboratory of Dark Matter Physics, Tsung-Dao Lee Institute, Shanghai Jiao Tong University, Shanghai 200240, People’s Republic of China}
\email{xzzheng@sjtu.edu.cn}

\author[orcid=0000-0001-7283-1100]{Xingchen Zhou}
\affiliation{National Astronomical Observatories, Chinese Academy of Sciences,20A Datun Road, Beijing 100012, People’s Republic of China}
\affiliation{Science Center for China Space Station Telescope, National Astronomical Observatories, Chinese Academy of Sciences,\\20A Datun Road, Beijing 100101, People’s Republic of China}
\email{xczhou@nao.cas.cn}

\author[orcid=0000-0002-6684-3997]{Hu Zou}
\affiliation{School of Astronomy and Space Sciences, University of Chinese Academy of Sciences(UCAS),\\Yuquan Road NO.19A Beijing 100049, People’s Republic of China}
\affiliation{Key Laboratory of Optical Astronomy, National Astronomical Observatories, Chinese Academy of Sciences, Beijing 100101, People’s Republic of China}
\email{zouhu@nao.cas.cn}

\correspondingauthor{Yingxiao Song, Xiaohu Yang}

\begin{abstract}

Because groups are directly associated with halos, they allow for considerably simpler theoretical modeling than approaches based on individual galaxies. We therefore propose to use voids identified in galaxy group catalogs, referred to as group-voids, to investigate the cosmic large-scale structure (LSS). Using the reference mock galaxy redshift survey (MGRS) designed for the Chinese Space-station Survey Telescope (CSST), we build two galaxy group catalogs representing ideal and realistic scenarios, derived from galaxy samples with 100\% and roughly 30\% spectroscopic redshift completeness, respectively. We then identify voids in these two mock group catalogs, as well as in the underlying halo catalog, and measure two void statistics, the void size function (VSF) and the void density profile, within five redshift intervals spanning $z=0$ to $1.0$. We compare the statistics obtained from two kinds of voids: those defined by galaxy groups (group-voids) and those defined by dark matter halos (halo-voids). In the void-finding process, we adopt the brightest central galaxy (BCG) as the group center to improve the accuracy of the inferred void centers. Our analysis shows that void statistics derived from group-voids with spectroscopic redshift completeness of at least 40\% can faithfully reproduce the corresponding statistics from halo-voids. Even when the redshift completeness of galaxies falls to as low as 30\%, we can still reliably describe group-voids via halo-voids by incorporating a redshift error term. This indicates that group-voids are a promising tool for probing LSS and offer a valuable complement to standard void studies, which is especially advantageous for emulator-based methods.

\end{abstract}


\section {introduction} \label{sec:intro}

Clusters, filaments, sheets and voids are the typical components of the cosmic large-scale structure (LSS). These structures, representing different cosmic web environments, have been proved in many existing studies to be powerful probes for exploring cosmology and galaxy evolution \citep[e.g.][]{2022ApJ...936..161W,contarini2023cosmological,2023Natur.619..269D,2024ApJ...969...89T,2025MNRAS.539.1692Z}. Among these components, the statistics of the cosmic voids, which evolve linearly due to their low density and large size, have emerged as a powerful way to put constraints on cosmological parameters \citep[e.g.][]{pisani2015counting,2022MNRAS.513..186A,2022A&A...658A..20H,2024A&A...691A..39R,2024MNRAS.534..128S,2024JCAP...10..079V,2025ApJ...993..227V}. 

However, the void samples in these works are all identified from galaxy catalogs. Research based on such voids generally needs the mock galaxy catalogs generated from high-precision simulations to calibrate nuisance parameters \citep[e.g.][]{2022A&A...667A.162C,contarini2023cosmological}, and the procedure for reconstructing the galaxy map from redshift space to real space is highly complicated. If we instead identify the void based on the galaxy groups, which are defined as sets of galaxies residing in the same dark matter halo, we can obtain the void samples from observational data which are similar to halo-voids, and we refer to these as group-voids. In principle, once galaxies are associated with halos, thereby eliminating the Finger-of-God (FoG) effect, it also becomes substantially easier to correct for the Kaiser effect during the reconstruction \citep[e.g.][]{2016ApJ...833..241S,2018ApJ...861..137S}. Most importantly, because groups are directly linked to halos, the theoretical modeling of group-void statistics can be performed directly using halo catalogs from numerical simulations. This enables the construction of accurate emulators, which are particularly powerful for cosmological parameter inference, a task that is far more complex and computationally expensive when using galaxy-based voids, which rely on empirical assumptions such as abundance matching or halo occupation distribution. Therefore, using group-voids to explore the LSS as a new probe will provide a novel perspective and complement traditional void studies, with the additional advantage of being naturally suited for emulator-based cosmological analyses.

Here, we propose to measure void statistics from void samples based on galaxy group catalogs for LSS analysis, including the void size function (VSF) and void density profile. The VSF is the function describing the number density distribution of voids by their size in a given redshift bin, which is the most widely studied and proved as an effective probe in the field of void research \citep[e.g.][]{2017MNRAS.469..787P,2019MNRAS.487.2836P,contarini2021cosmic,2023MNRAS.522..152P,2024MNRAS.532.1049S,2024JCAP...10..079V,2025MNRAS.540.2853S}. While the void density profile denotes the mean deviation in spherical shells between the void density and the background mean density. They were commonly employed to determine the correlation function and power spectrum of voids \citep[e.g.][]{2016MNRAS.462.2465C,2017MNRAS.465..746S,2018MNRAS.476.3195C,2019MNRAS.483.3472N,2021MNRAS.500..464V,2022MNRAS.516.4307W,2023A&A...670A..47B,2023JCAP...08..010V,2023A&A...674A.185M,2023A&A...677A..78R,2024ApJ...976..244S,2025MNRAS.538..114S}. Hence, in this work, we will measure these two void statistics in group-void samples and halo-void samples, and discuss the potential of group-voids, particularly their suitability for emulator-based modeling. 

\begin{figure*}
\subfigure{
\includegraphics[width=\columnwidth]{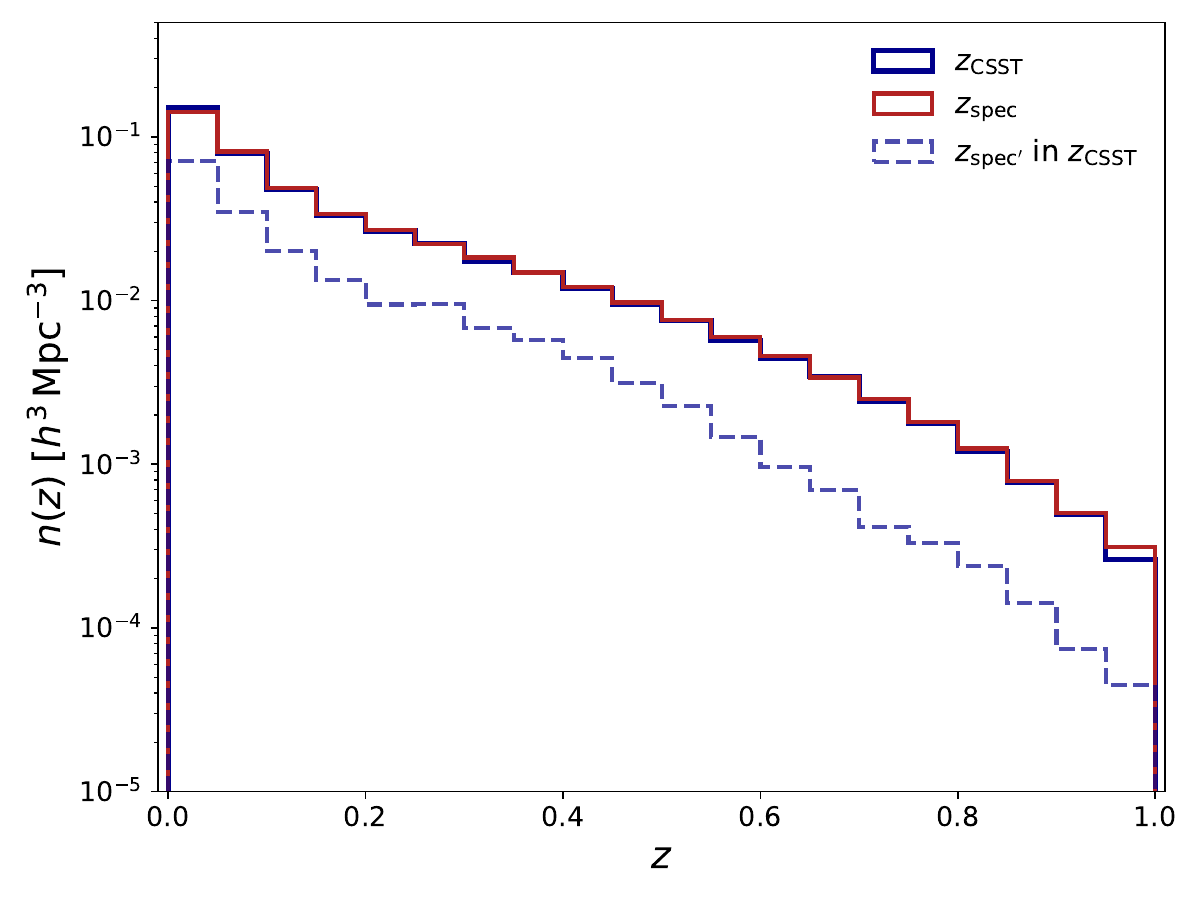}}
\hspace{2mm}
\subfigure{
\includegraphics[width=\columnwidth]{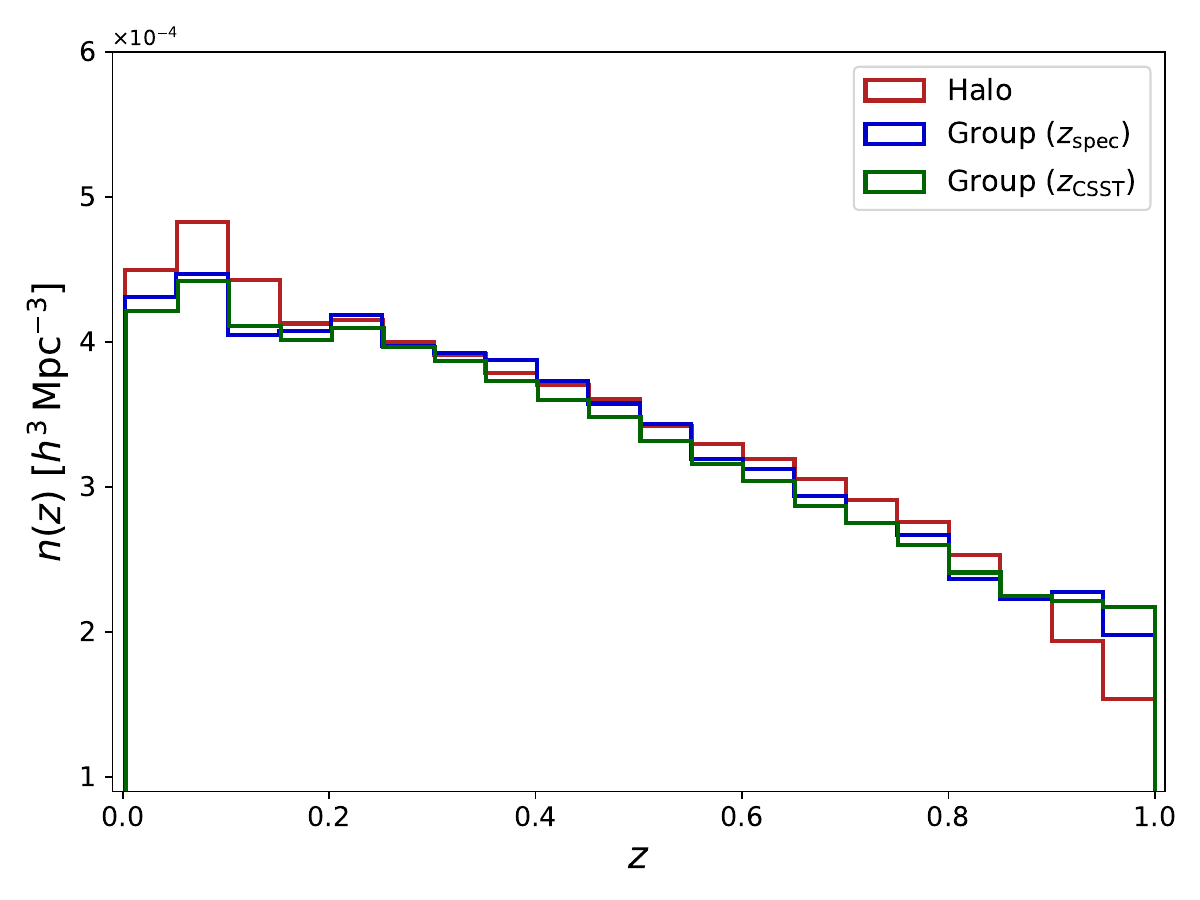}}
\caption{Left panel: The galaxy number density distribution from the ideal case ($z_{\rm spec}$) and the realistic case ($z_{\rm CSST}$) at $z<1$. The dashed line shows the number distribution of galaxies with spectroscopic redshifts ($z_{\rm spec'}$) in the realistic case.
Right panel: The number density distribution of halos and of groups from $z_{\rm spec}$ and $z_{\rm CSST}$ with $\log [M / h^{-1}M_{\odot}]\geq 13$ at $z=0-1$.  }\label{fig:ng}
\end{figure*}

To assess the reliability of this method, and to establish a foundation for emulator-based cosmological modeling using halo catalogs, first we identify the group samples by an extended halo-based group finder \citep{2005MNRAS.356.1293Y,2007ApJ...671..153Y,2012ApJ...752...41Y,2021ApJ...909..143Y} from the mock galaxy catalog, which is constructed according to the spectroscopic survey strategy and instrumental design of the Chinese Space-station Survey Telescope \citep[CSST,][]{zhan2021csst,2023MNRAS.519.1132M,gong,2025SCPMA..6880402G,2026SCPMA..6939501C}. Then we use the void finder, based on the Voronoi tessellation and the watershed algorithm to identify void samples from group catalogs and halo catalogs. Next, we measure the VSF and void density profile in five redshift bins at $z=0.0-1.0$, and compare the differences of these two void statistics from halo-voids and group-voids in ideal and realistic cases. We further include a free parameter $\sigma_0$ to account for the influence of the CSST photo-z effect arising from the absence of reliable slitless spectroscopic redshift measurements on the group VSFs.

The paper is organized as follows: In Section \ref{sec:data}, we introduce the mock galaxy catalog, the group finder and the void finder we use; In Section \ref{sec:vs}, we discuss the measurements of the VSF and void density profile; The summary and conclusion are given in Section \ref{sec:conclusion}.

\section{MOCK DATA} \label{sec:data}

\subsection{Galaxy Catalog} \label{sec:galcat}

We use the reference mock galaxy redshift survey \citep[MGRS,][]{2024MNRAS.529.4015G} for CSST to proceed with the identification of group and group-void samples. 
The MGRS is a mock catalog designed to simulate the CSST spectroscopic survey, generated from a high-resolution N-body simulation ``Jiutian" \citep{2025SCPMA..6809511H}.
This simulation uses a box of side $1\,h^{-1}\mathrm{Gpc}$, contains $6144^3$ particles, and has a mass resolution of $m_{\mathrm{p}} = 3.723\times10^8\,h^{-1}M_\odot$. It adopts the cosmological parameters from $\it Planck$2018 \citep{2020A&A...641A...6P}, i.e. $\Omega_\Lambda = 0.6899$, $\Omega_\text{m} = 0.3111$,  $\Omega_\text{b} = 0.0490$, $n_\text{s} = 0.9665$, $\sigma_8 = 0.8102$, and $h = 0.6766$. We use the subhalo abundance matching (SHAM) method to obtain the galaxy luminosities for all subhalos according to the luminosity function at z-band for the Dark Energy Spectroscopic Instrument \citep[DESI,][]{2016arXiv161100036D}. 
To assign realistic galaxy properties, we match the galaxy samples to the group catalog of the DESI Legacy Imaging Surveys Data Release 9 (LS DR9) using luminosity, halo mass and redshift to obtain the color, stellar mass, star-formation rate and even image for each galaxy.
Overall, the mock galaxy catalogs have a magnitude cut of less than 21 in z-band and only consider sources at $z = 0 - 1.0$ based on DESI LS DR9 data. This MGRS can represent the ideal case for CSST slitless spectroscopic survey, i.e., all galaxies in this catalog have spectroscopic redshift measurements, and it covers $\sim$18350 square degrees of the sky and contains more than 118 million mock galaxy samples with cosmological redshifts, as well as spectroscopic redshifts taking into account redshift space distortion (RSD) effect and a typical redshift error $35\ {\rm km\ s}^{-1}$ in the current spectroscopic redshift surveys at $z < 1.0$. We will use $z_{\rm spec}$ to represent the spectroscopic redshift values in this catalog, which include both peculiar motions and cosmological redshift, and we identify group samples using these redshifts as the ideal case in our analysis.

However, in the actual CSST survey, slitless spectroscopy will not provide us with such a large number of sources with measured redshifts, so we must take the related observing conditions into account when assessing the robustness of our method. Therefore, we employ the CSST Emulator for Slitless Spectroscopy \citep[CESS,][]{2024MNRAS.528.2770W} on MGRS to generate the mock slitless spectra, and estimate the mock magnitudes for each band based on galaxy spectral energy distributions (SEDs) according to all filters of the CSST camera. Then we measure the redshift from these slitless spectra \citep[e.g.][]{2024ApJ...977...69Z,2025MNRAS.538..395S}, and to achieve higher completeness, we choose an XGBoost-based classifier that uses spectroscopic diagnostics and photometric properties to check the quality of measured redshifts \citep{2026ApJ..1003..137P}. The slitless spectroscopic redshifts flagged as reliable by the XGBoost-based classifier exhibit an accuracy exceeding 95\%. Here, accuracy is defined as $\Delta z \leq 0.002 (1 + z)$. These high-quality measurements constitute more than 20\% of the sources listed in the original catalog. 
Then we match the samples that have accurate redshifts with the MGRS, thus resulting in $\sim$26 million galaxy samples with slitless spectroscopic redshift at $z = 0 - 1.0$. Note that we also match the redshift results from DESI Year 1 (Y1) Bright Galaxy Survey (BGS) data to maximize completeness, since our mock galaxy samples were generated based on DESI LS DR9 data, and the accepted criterion includes 100\% of samples with z-band magnitude $m_{\rm z}\leq 17.7$ and $\sim50\%$ of samples with $17.7<m_{\rm z}\leq 19.5$. In general, more than 30\% of the galaxies in this final catalog have redshifts that achieve the CSST spectral calibration accuracy of $\Delta{z}/(1 + z) \leq 0.002$, with $\sim38$ million sources. We refer to this combined set from CESS results and DESI Y1 BGS spectroscopic redshifts as $z_{\rm spec'}$. For the other galaxy samples, we assign a new redshift using a Gaussian distribution based on a photometric redshift error with $\sigma_{\rm z} = (0.01 + 0.015z_{\rm spec})(1 + z_{\rm spec})$, which reflects the photometric redshift uncertainty in a real survey. 
Finally, we denote the resulting redshift assigned to each galaxy in this new catalog as $z_{\rm CSST}$, which represents the realistic case based on our simulation for the CSST survey.

Thus our analysis will be based on two redshift catalogs constructed from the same MGRS. The ideal catalog provides a $z_{\rm spec}$ value for every galaxy in MGRS; the realistic catalog assigns instead $z_{\rm CSST}$, which equals $z_{\rm spec'}$ for galaxies that meet the CSST spectroscopic accuracy requirement, and a photometric redshift estimate otherwise. The total number of galaxies is $\sim$118 million in the ideal case and $\sim$116 million in the realistic case, with the small difference ($<$ 2\%) caused by the matching process. Because the redshift coordinates differ, the galaxies falling into a given $z$-bin in the two catalogs are generally not the same. Nevertheless, the galaxy number density distributions of the two catalogs are similar, and we find that the number density values are about $\bar n =  4.5\times 10^{-2}, 1.8\times 10^{-2}, 8.3\times 10^{-3}, 2.9\times 10^{-3}$, and $6.6\times 10^{-4}$  $h^3{\rm Mpc}^{-3}$ for the five redshift bins from $z=0$ to 1.0 with a bin width of 0.2, respectively. 
In the left panel of Figure~\ref{fig:ng}, we show the galaxy number density distribution of the two catalogs, and the number distribution of spectroscopic redshifts from the realistic case is also shown.

\subsection{Group Catalog} \label{sec:groupcat}

We use the extended halo-based group finder \citep{2021ApJ...909..143Y} to construct the group catalogs from our two mock galaxy catalogs. Our group finder can identify group samples from galaxy samples with spectroscopic and photometric redshifts by accounting for different redshift errors. The steps involved in our group finder are as follows:

\begin{itemize}
 \item Consider each galaxy as a group candidate initially.
 \item Measure the accumulative luminosity functions of the groups for different redshift bins.
 \item Determine the mass-to-light ratio for groups in each redshift bin by abundance matching to an accumulative halo mass function assuming a given cosmology. 
 \item Assign each tentative group with the mass, size, and velocity dispersion of the halo. 
 \item Search member galaxies of each group candidate using its associated halo information.
 \item Iterate until the group memberships become stable and the mass-to-light ratios reach convergence. 
\end{itemize}
The group samples identified by our group finder provide several important group properties, such as the luminosity-weighted center, mass, galaxy memberships, and the brightest central galaxy (BCG) in each group sample. 
The luminosity-weighted center is computed by averaging the coordinates of member galaxies, with each galaxy weighted by its luminosity. In our analysis, we consider two types of group centers: the luminosity-weighted center and the BCG position. We note that the group-void properties obtained using these two types of centers are largely consistent. Consequently, in the following analysis we use, as a representative example, the void statistics based on group-voids identified with the luminosity-weighted center, provided no significant differences arise between the two definitions.

To ensure the completeness of the group samples, we focus on massive groups with $\log [M / h^{-1}M_{\odot}]\geq 13$ in our analysis and we find that both group catalogs contain more than 6 million groups in $z = 0 - 1.0$. The number density distributions of groups identified from the two galaxy catalogs are shown in the right panel of Figure~\ref{fig:ng}, and we also show the number density distribution of halos with $\log [M / h^{-1}M_{\odot}]\geq 13$ from our simulation as a comparison. Note that the halo redshifts used in our analysis also include the RSD effect. The results indicate that the number density distributions of our group catalogs are consistent with that of the halo catalog. We also compute the mean spacing between groups and halos as $\bar n^{-1/3}$ in different redshift bins, and for two group and one halo catalogs across the five redshift bins, the values are consistently close to 15 $h^{-1}\text{Mpc}$.

\begin{figure*}
\centering
\includegraphics[width=2\columnwidth]{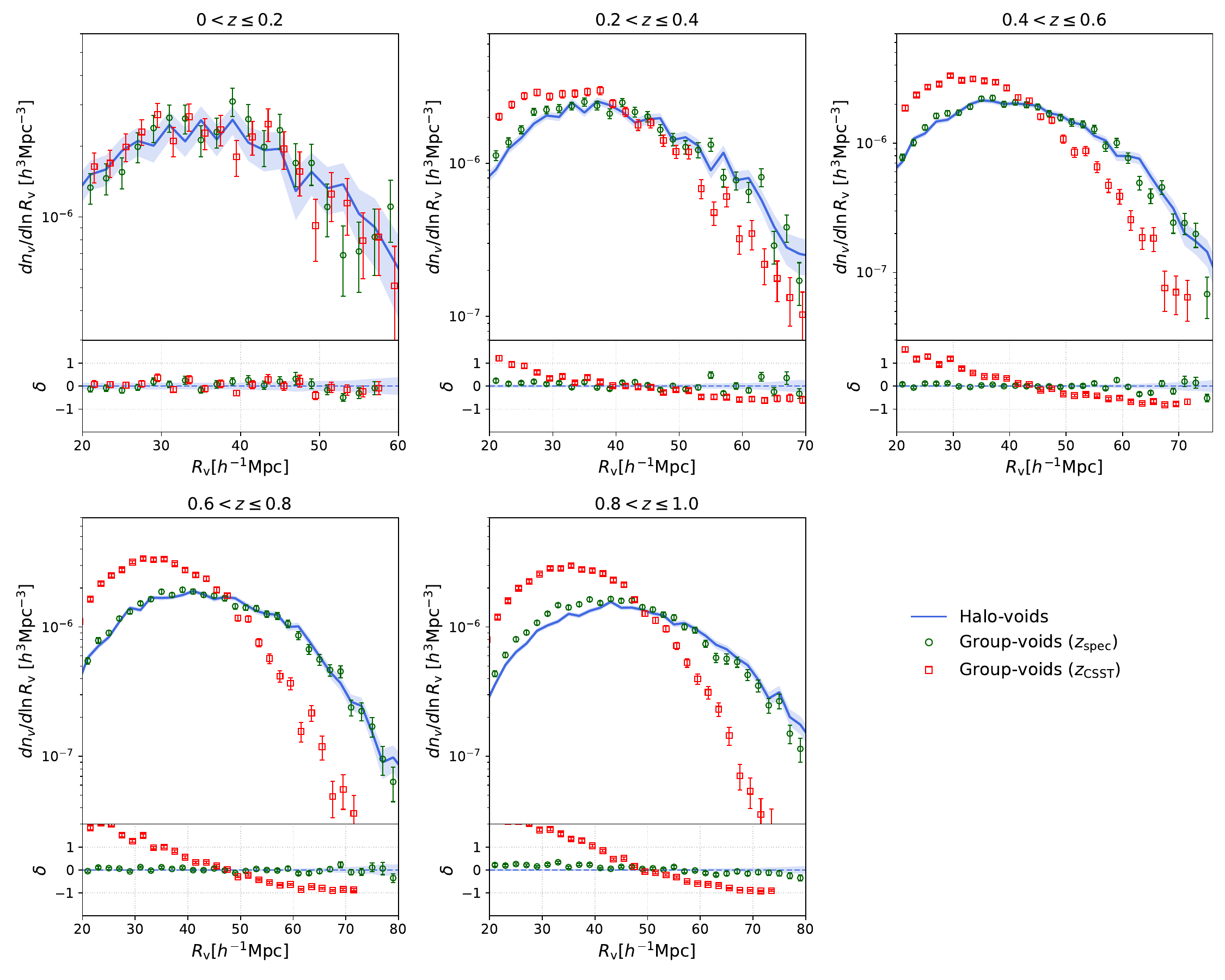}

\caption{The VSF data from halo catalog (blue) and group catalogs with $z_{\rm spec}$ (green) and $z_{\rm CSST}$ (red) in the five redshift bins. The small subpanel shows the relative deviation $\delta$ between the VSF of the two group-void samples and that of the halo-voids. The shaded regions and error bars indicate the $1\sigma$ uncertainty of the VSF data.}
\label{fig:vsf}
\end{figure*}

\begin{deluxetable}{ccccc}
\tablenum{1}
\tablecaption{The void numbers $N_{\rm hv}$ from the halo catalog, $N_{\rm gv}^{z_{\rm spec}}$ from the group catalog with $z_{\rm spec}$, and $N_{\rm gv}^{z_{\rm CSST}}$ from the group catalog with $z_{\rm CSST}$ in the five redshift bins.}\label{tab:cat}
\setlength{\tabcolsep}{12pt}  
\tablewidth{\textwidth} 
\tablehead{
\colhead{$z_{\rm min}$} & \colhead{$z_{\rm max}$} &\colhead{$N_{\rm hv}$ } &\colhead{$N_{\rm gv}^{z_{\rm spec}}$}&\colhead{$N_{\rm gv}^{z_{\rm CSST}}$} 
}
\startdata
    0&0.2&850&848&868\\
    0.2&0.4&4478&4651&5746\\
    0.4&0.6&8559&8597&12452\\
    0.6&0.8&11503&11541&18904\\
    0.8&1.0&11922&13476&21452\\
\enddata
\end{deluxetable}

\subsection{Void Catalog} \label{sec:voidcat}
\setcounter{footnote}{0}
We generate void catalogs from our halo catalog and two group catalogs using the void finder \texttt{VIDE} \citep[][]{vide} from The Void Analysis Software Toolkit\footnote{\url{https://github.com/desi-ur/vast}} \citep[\texttt{VAST},][]{vast}. 
The method applies Voronoi tessellation in combination with the watershed algorithm \citep{watershed} to detect underdense regions, following the framework of the ZOnes Bordering On Voidness method \citep[\texttt{ZOBOV},][]{zobov}. A key advantage of voids identified with \texttt{VIDE} is that no specific shape is assumed, and the toolkit provides various void properties including volume-weighted center and effective radius. Although \texttt{VIDE} can merge adjacent underdensities based on the tracer number density at their boundaries, we opt to use the unmerged zones in our analysis to avoid the void-in-void issue.

The void volume $V$ and effective radius $R_{\rm v}$ are derived from the Voronoi cells associated with each tracer. By summing the volumes of the cells that comprise a void, its total volume can be derived as $V = \sum_{i=1}^N V_{\rm cell}^i$, where $V_{\rm cell}^i$ is the volume of the $i$-th cell and $N$ is the number of cells within the void. Based on this total volume, the corresponding effective radius $R_{\rm v}$ under the assumption of a spherical approximation, is then computed as $R_{\rm v} = (3V/4\pi )^{1/3}$. In addition to the size, we also characterize the position of each void by its volume-weighted center $\mathbf{X}_{\rm v}$, which is obtained by averaging the tracer positions $\mathbf{x}_i$ weighted by their respective Voronoi cell volumes as $\mathbf{X}_{\rm v} = \sum^N_{i=1}\mathbf{x}_i V^i_{\rm cell} / V$.

\begin{figure*}
\centering
\includegraphics[width=1.8\columnwidth]{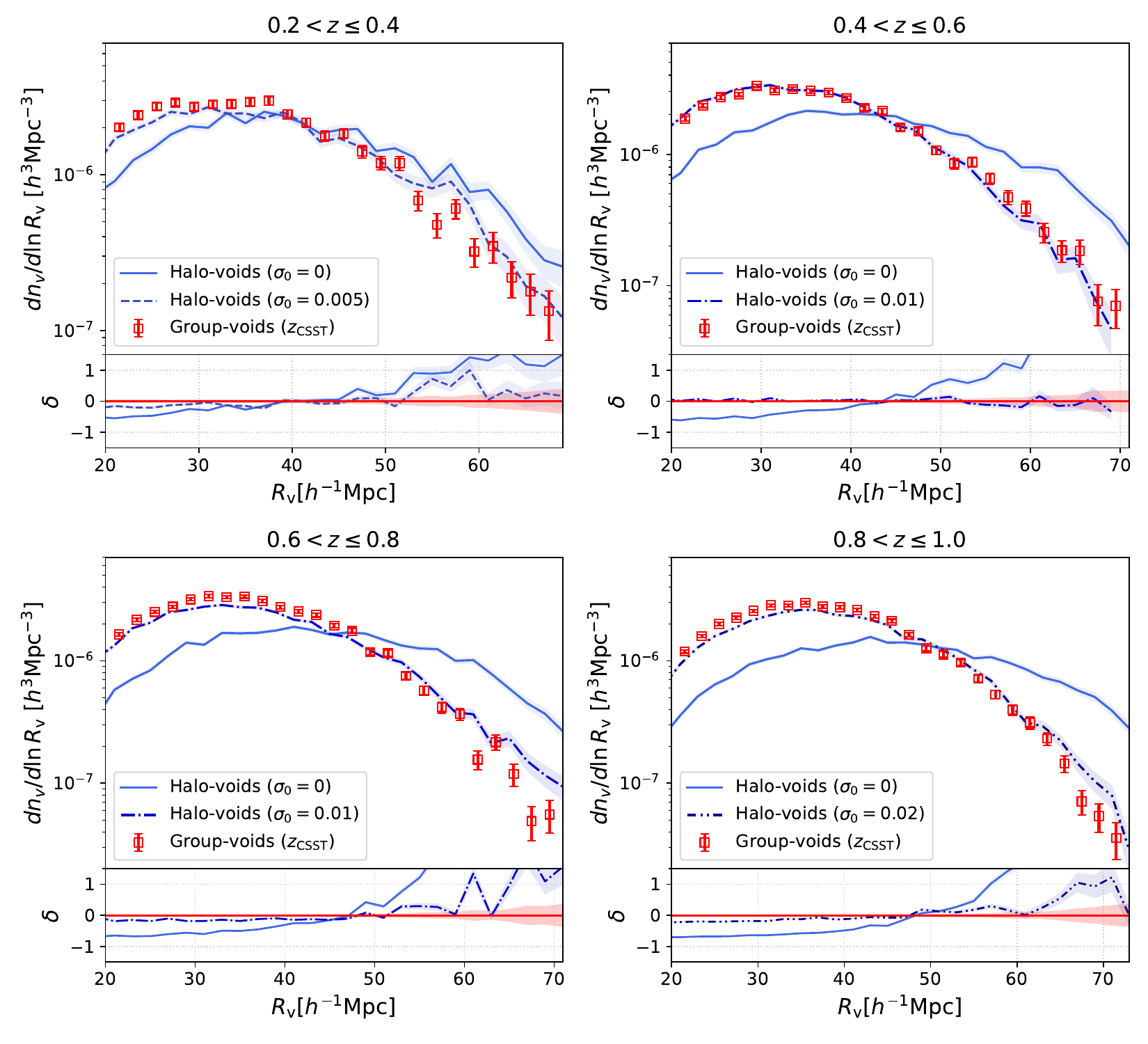}

\caption{The VSF data from the halo catalogs (blue) and group catalogs with $z_{\rm CSST}$ (red) in four redshift bins. The blue color becomes darker and the line style changes from solid to dashed, dash-dotted, and dash-double-dotted with increasing additional redshift errors applied to the halo catalog ($\sigma_0 = 0, 0.005, 0.01, 0.02$). The small subpanel shows the relative deviation $\delta$ between the VSF of the halo-voids with additional redshift errors and that of the halo-voids with $\sigma_0 = 0$ in each redshift bin. The shaded regions and error bars indicate the $1\sigma$ uncertainty of the VSF data.}
\label{fig:hvsf}
\end{figure*}

In Table~\ref{tab:cat}, we show the number of voids $N_{\rm v}$ for our three void catalogs. We find that the void number is consistent between the halo-void and group-void from group catalogs with $z_{\rm spec}$, and that there is no order of magnitude difference in the number of group-voids from the two group catalogs. In addition, we notice that the void minimum radius in our three catalogs is similar at different redshift bins, approximately 5 $h^{-1}\text{Mpc}$. The values of the maximum radius from the first redshift bin to the last redshift bin are about 75 - 100 $h^{-1}\text{Mpc}$ for the group-void samples with $z_{\rm spec}$ and the halo-void samples. And for the group-void samples with $z_{\rm CSST}$, the maximum radius values are roughly similar in the five redshift bins, with values around 80 $h^{-1}\text{Mpc}$.

\section{VOID STATISTIC}\label{sec:vs}

In this section, we describe the measured results of two void statistics, including the void size function and the void density profile, and compare the results from our two group catalogs and the halo catalog to investigate the feasibility of group-voids.

\subsection{Void Size Function} 
\label{sec:vsf}

In Figure~\ref{fig:vsf}, we show the void size function data, ${\rm d}n_{\rm v}/{\rm d\,ln}R_{\rm v}$, for the halo-voids, group-voids from $z_{\rm spec}$ and $z_{\rm CSST}$ in the five redshift bins. We choose the VSF from halo-voids as a reference to evaluate the accuracy of the VSF data from group-voids by calculating the relative deviation $\delta$, and we use the jackknife method \citep{2021MNRAS.501.3309Z} to estimate the $1\sigma$ error.
Note that we focus on the VSF data of voids with $R_{\rm v}>20\ h^{-1}\text{Mpc}$ to ensure sufficient completeness of the void samples, a threshold that is approximately 1.5 times the mean separation of groups and halos in all redshift bins. Additionally, void radius bins containing fewer than five voids are excluded for statistical significance.

We find that the VSFs from group-voids of $z_{\rm spec}$ are consistent with the results from halo-voids within $1\sigma$ error at $z<0.8$. And for the VSF at $0.8 < z \leq1.0$, the results for large-size void samples with radius $\gtrsim40 h^{-1}\text{Mpc}$ are also consistent. These results indicate that the VSF of group-voids from galaxy samples with full spectroscopic redshifts can represent the distribution of halo-void sizes.

However, we also find that the results from group-voids of $z_{\rm CSST}$ are generally not consistent with the VSF of halo-voids, except in the first redshift bin. Nevertheless, for voids with sizes $\gtrsim40 h^{-1}\text{Mpc}$ at $0.2 < z \leq0.4$, we notice the differences are not significant, with mean deviations of less than 2$\sigma$. This result can be attributed to the varying fractions of galaxies with spectroscopic redshifts across different redshift bins. The fractions of galaxies with $z_{\rm spec}$ in the galaxy catalog of $z_{\rm CSST}$ for the first two redshift bins ($z<0.4$) are approximately 40\%, and there are $\sim30\%$ for the redshift bin at $0.4 < z \leq0.6$, while they drop to $\sim20\%$ in the last two redshift bins ($z>0.6$). The higher spectroscopic completeness at lower redshift range contributes to the better performance of voids in these redshift bins, especially for large voids. Indeed, previous studies on cosmological constraints using the VSF data usually exclude small voids \citep[e.g.][]{2019MNRAS.488.5075R,2019MNRAS.488.3526C,contarini2021cosmic,2022A&A...667A.162C,contarini2023cosmological,2024A&A...682A..20C,2025MNRAS.540.2853S}. 

Based on our VSF results from group-voids of $z_{\rm CSST}$, we find that group-voids can tolerate a certain degree of spectroscopic incompleteness, in contrast to voids identified from galaxy samples in previous studies, which generally require full spectroscopic redshifts \citep[e.g.][]{contarini2023cosmological,2024ApJ...969...89T,2025MNRAS.540.2853S}. While inaccurate tracer positions strongly affect void sizes, the suppression in void size caused by spectroscopic incompleteness is weaker in group-void samples. This is because group samples are defined by sets of member galaxies, making them less sensitive to individual redshift errors. 
When groups are used as tracers to identify voids, the group position is more robust than that of individual galaxies. Hence, the void position derived from group tracers is more stable than that identified on any single member galaxy.

Note that we compare the void number and VSFs of the group-voids from our two cases identified with the two kinds of centers (luminosity-weighted center and BCG) and find that the results are consistent at different redshift bins. This indicates that using different group center does not significantly affect the void abundance or size distribution. 

In contrast, traditional VSF analysis is performed on galaxy samples, and it usually needs related mock galaxy catalogs to calibrate nuisance parameters \citep[e.g.][]{2022A&A...667A.162C,contarini2023cosmological}. Although there are several approaches to placing galaxies into mock halo catalogs, including abundance matching, the halo occupation distribution, and semi-analytic models \citep[e.g.][]{2018ARA&A..56..435W}, these methods either rely on empirical assumptions or require complex modeling with high computational cost. However, if we employ the VSF derived from group-voids to constrain cosmological parameters, we can work directly with halo-voids in simulations and thereby greatly simplify the calibration of nuisance parameters compared to galaxy-void measurements. This approach is expected to result in a more streamlined analysis and improve the precision of cosmological constraints.

Moreover, even for the VSF derived from group-voids constructed with partial spectroscopic redshifts, which are affected by the photometric redshift errors illustrated in Figure~\ref{fig:vsf} for $z_{\rm CSST}$, it is straightforward to capture this effect by introducing a free parameter that characterizes the halo redshift error. Concretely, we assign each halo in the simulation catalog a new redshift drawn from a Gaussian distribution that incorporates a redshift uncertainty, with $\sigma_{\rm z} = \sigma_0(1 + z)$, and then investigate how this modifies the VSF of these halo-voids.

We apply redshift errors with $\sigma_0 = 0.005$, $0.01$, and $0.02$ to the halo catalog, and compare the VSFs of these halo-voids with that of the group-voids from $z_{\rm CSST}$ at $z>0.2$ in different redshift bins. 
As shown in Figure~\ref{fig:hvsf}, compared to the case without redshift errors ($\sigma_0 = 0$), the VSFs of halo-voids with $\sigma_0 = 0.005$, $0.01$, and $0.02$ show substantially reduced deviations from the results of group-voids ($z_{\rm CSST}$) in the redshift ranges $0.2<z\leq0.4$, $0.4<z\leq0.8$, and $0.8<z\leq1.0$, respectively. This suggests that, for group-voids affected by spectroscopic incompleteness and the resulting photometric redshift uncertainties, the impact of these effects can be largely mitigated by a single parameter within the halo-void framework. By allowing $\sigma_0$ to vary as a free parameter, we can fit the VSF and thus constrain its value.
With these considerations, the group void measurements become well-suited for cosmological analyses. We plan to explore this approach in future work.

\begin{figure*}
\centering
\includegraphics[width=2\columnwidth]{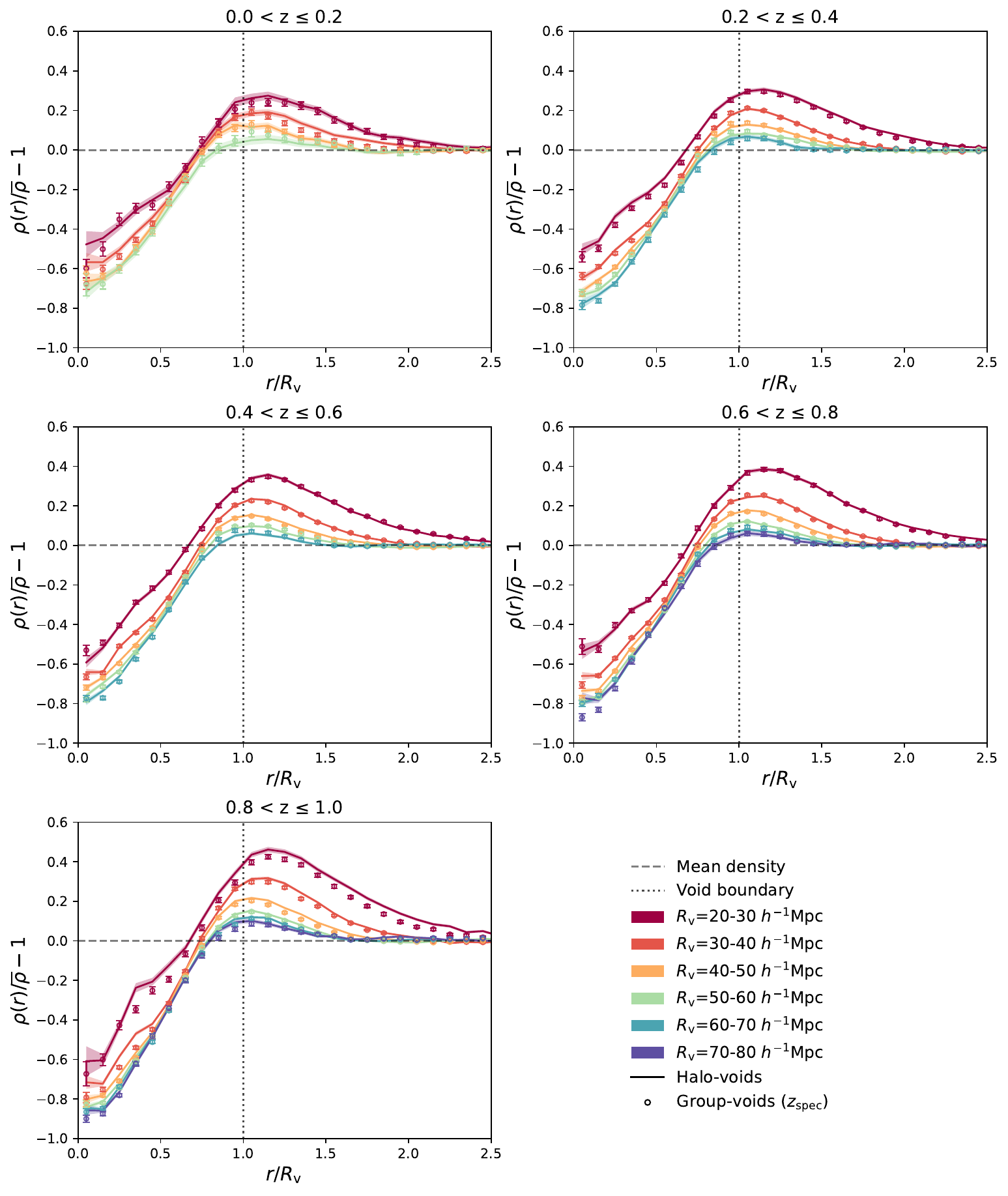}

\caption{Stacked void density profile from the halo catalog and the group catalogs with $z_{\rm spec}$ in the five redshift bins. Colors from red to purple indicate the void radius bins ranging from $R_{\rm v}=20$ to 80 $h^{-1}\text{Mpc}$. The shaded regions and error bars represent the $1\sigma$ error. The horizontal dashed line and vertical dotted line denote the mean background density and the void boundary ($r/R_{\rm v}=1$), respectively.}
\label{fig:gvdp1}
\end{figure*}

\subsection{Void Density Profile} \label{sec:vdp}

To further check the reliability of group-voids, we also measure the stacked void density profile of group-voids and halo-voids at different redshift bins, which is defined as $\rho_{\rm v}(r)/ \bar{\rho}-1$, here $\rho_{\rm v}(r)$ is the void density and $\bar{\rho}$ is the mean density of the universe. 
Note that here the density profile measures the galaxy number density around the void center. We estimate this profile using the same galaxy sample with $z_{\rm spec}$ (i.e., the MGRS catalog with 100\% spectroscopic redshifts) for all void samples, including the group-voids from both the ideal and realistic cases, as well as the halo-voids. This allows us to assess whether the void centers identified from the two group-void samples are consistent with the halo-void positions. Moreover, using the galaxy sample to measure the density profile is not influenced by potential biases from the group-finding process and enables a clean comparison with the halo-void results.

In Figure~\ref{fig:gvdp1}, we show the stacked void density profile for the halo-voids and group-voids from the galaxy catalog with $z_{\rm spec}$ in the five redshift bins, and we use the standard error of the mean profile, $\sigma/\sqrt{N_{\rm v}}$, as the statistical uncertainty, where $\sigma$ is the standard deviation of the individual void profiles within the stack and $N_{\rm v}$ is the number of voids in the stack. Note that we use the radius of each void to normalize the density profile during measurement. We adopt selection criteria similar to those in the VSF analysis, focusing on voids with $R_{\rm v}>20\ h^{-1}\text{Mpc}$, and we choose the upper limits of the largest sizes of voids for the five redshift bins as $\sim60,70,70,80,80h^{-1}\text{Mpc}$ for statistical significance, according to the number of voids in each radius bin. Note that the radius bins used in this analysis have a width of $10h^{-1}\text{Mpc}$ across all redshift bins.

We notice that the density profiles from group-voids of $z_{\rm spec}$ are consistent with the results from halo-voids within $1\sigma$ in almost all radius bins at $z=0-1$. These results indicate that group-voids from galaxy samples with full spectroscopic redshifts can reflect the position and depth of halo-voids.

\begin{figure*}
\centering
\includegraphics[width=2\columnwidth]{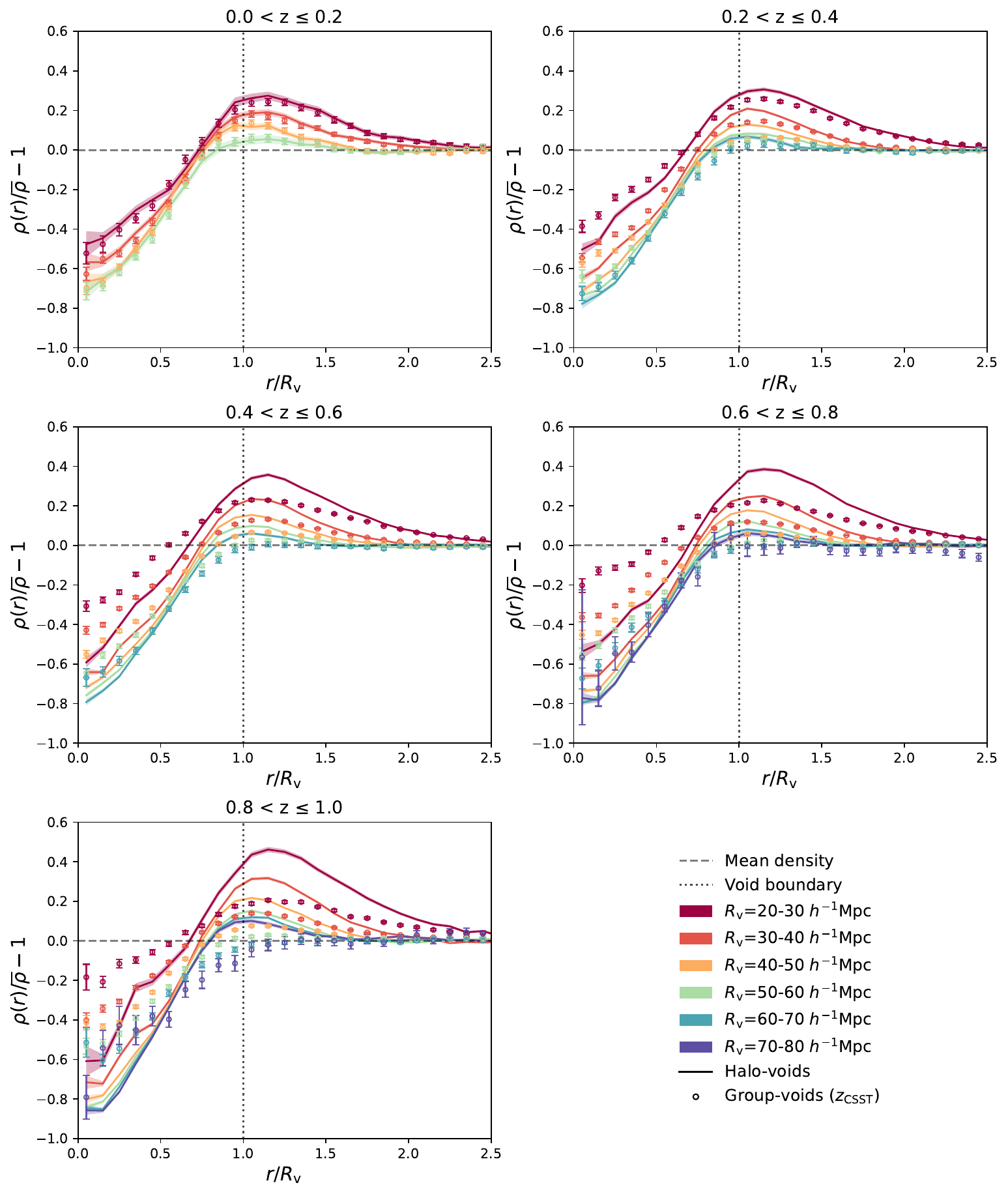}

\caption{Stacked void density profile from the halo catalog and the group catalogs with $z_{\rm CSST}$ in the five redshift bins. Colors from red to purple indicate the void radius bins ranging from $R_{\rm v}=20$ to 80 $h^{-1}\text{Mpc}$. The shaded regions and error bars represent the $1\sigma$ error. The horizontal dashed line and vertical dotted line denote the mean background density and the void boundary ($r/R_{\rm v}=1$), respectively.}
\label{fig:cgvdp}
\end{figure*}

In Figure~\ref{fig:cgvdp}, we show the stacked void density profiles from the halo-voids, as well as from the group-voids with $z_{\rm CSST}$ using the luminosity-weighted center, at $z\leq1.0$. We find significant deviations of the group-void density profiles relative to the halo-void results at $z >0.2$, especially for samples with $z>0.6$. These deviations are expected because the fraction of galaxies with spectroscopic redshifts ($z_{\rm spec'}$) in the galaxy catalog with $z_{\rm CSST}$ decreases with redshift, reducing the accuracy of the luminosity-weighted center for groups.

\begin{figure*}
\centering
\includegraphics[width=2\columnwidth]{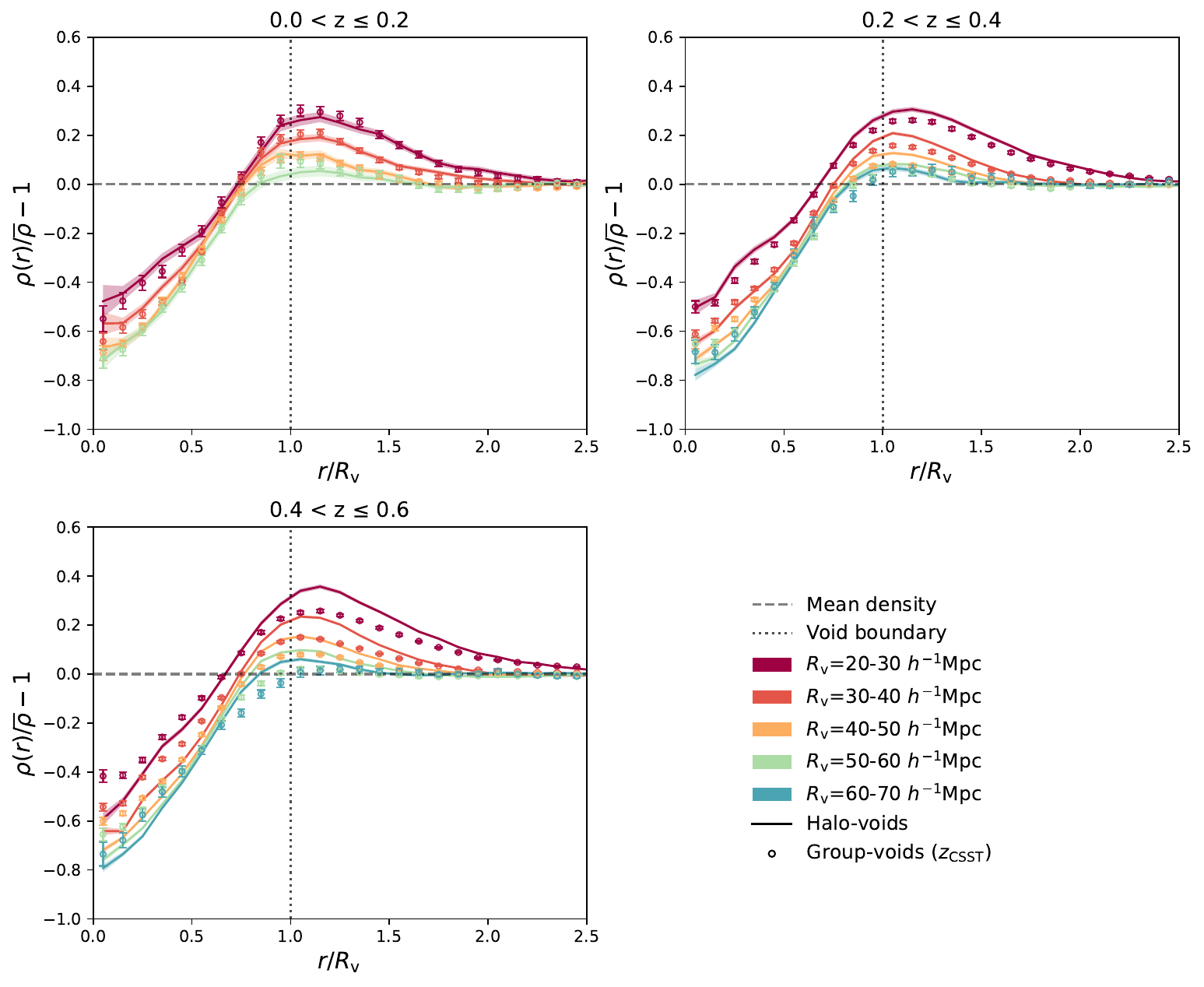}

\caption{Stacked void density profile from the halo catalog and the group catalogs with $z_{\rm CSST}$ identified using BCG in the first three redshift bins. Colors from red to cyan indicate the void radius bins ranging from $R_{\rm v} =20$ to 70 $h^{-1}\text{Mpc}$. The shaded regions and error bars represent the $1\sigma$ error. The horizontal dashed line and vertical dotted line denote the mean background density and the void boundary ($r/R_{\rm v}=1$), respectively.}
\label{fig:gvdp2}
\end{figure*}

Nevertheless, we find an improvement in the void positions when using BCG as the tracer for group-void identification, since we consider that the fraction of spectroscopic redshifts among BCGs would be higher.
In Figure~\ref{fig:gvdp2}, we show the void density profile for the halo-voids and group-voids of $z_{\rm CSST}$ identified based on BCG at $z\leq0.6$. We find that the density profiles around the void center from group-voids of $z_{\rm CSST}$ are almost consistent with the results from halo-voids at $z=0-0.6$. We notice that the improvement is most significant in the regions with $r/R_{\rm v} < 0.8$, where the density profiles from BCG-based group-voids show better agreement with the results from halo-voids. This means group-voids identified using BCG can improve the accuracy of locating low-density regions.
We find in the inner regions of the voids ($r/R_{\rm v} < 0.8$), the density profiles obtained with BCG as the group center instead of the luminosity-weighted center exhibit reduced average deviations from the halo-void results across the radius bins, by approximately 18.3\%, 40.8\%, and 43.4\% in the first three redshift bins, respectively.

For the results near the void boundary, we notice that there is only a slight improvement. This is because the incomplete redshift information in the group catalog with $z_{\rm CSST}$ distorts the void size distribution. As shown in the VSF results (Figure~\ref{fig:vsf}), large voids appear squeezed to smaller sizes and small-radius bins are filled with incomplete voids. Since the effective radius $R_{\rm v}$ reflects the void boundary, this distortion directly degrades the accuracy of void boundary determination. 
However, we do not show the density profiles from BCG-based group-voids for $z=0.6-1.0$, since the fraction of galaxies with $z_{\rm spec}$ in this redshift range is too low ($\sim20\%$) to achieve significant improvement. 
Therefore we consider the improvement from using BCG as the group center to be significant in the inner regions of the voids, though it remains limited near the void boundary.

In fact, previous void analyses based on galaxy samples identify voids directly in redshift space. Correcting for the RSD effect in such analyses typically requires introducing nuisance parameters in the modeling or performing complex reconstructions\citep[e.g.][]{correa2021redshift,2019PhRvD.100b3504N}, and the reconstruction for galaxy samples from redshift space to real space is complicated by both the FoG and Kaiser effects, requiring additional assumptions. However, once galaxies are associated with their host groups, the FoG effect is largely mitigated because the random peculiar motions of member galaxies are averaged out. Moreover, group centers offer more stable and trustworthy reference positions than individual galaxies. Working with groups also makes it easier to correct for the Kaiser effect, because the large-scale velocity field can be reconstructed directly from the group catalog itself \citep[e.g.][]{2016ApJ...833..241S,2018ApJ...861..137S}. Conversely, in simulation-based emulators one can straightforwardly include the halo velocities to account for this effect.  Thus, group-voids lead to a more accurate determination of void positions and density profiles under realistic conditions (e.g., the realistic case in our work), consistent with our findings that group-void sizes are less affected by redshift measurement uncertainties.  In future cosmological analyses, we can avoid the influence of nuisance parameters related to RSD effects, because the reconstruction for group samples is simpler and more accurate.

\section{CONCLUSION}
\label{sec:conclusion}

In this work, we investigate the potential of group-voids for exploring the LSS by measuring the VSF and void density profile. We identify the group samples from the two galaxy catalogs with different fractions of accurate redshifts based on the CSST MGRSs, and identify voids from the halo catalogs and these two group catalogs. The VSFs and density profiles are then derived based on the group-voids and halo-voids at the five redshift bins from $z=0$ to $1.0$. We set the results of VSF and density profile from halo-voids as a reference at a given redshift. Then we compare the measured results from two kinds of group-voids and halo-voids, and analyze the deviations.

We find that, in the ideal case where group-voids are derived from the galaxy catalog with full accurate redshifts, the VSF and density profile measurements are consistent with those from halo-voids. In contrast, in the realistic case where group-voids come from the galaxy samples with only partial accurate redshifts, the measurements exhibit significant deviations at $z>0.6$, where the fraction of spectroscopic redshifts drops below 30\%. This is because the accuracy of tracer positions affects void identification, and the resulting void properties directly influence the measured void statistics, leading to the observed deviations. We also find that using the BCG as the group center in void identification can effectively improve the accuracy of group-void centers, but this method is effective only when the fraction of spectroscopic redshifts exceeds 30\% in a given redshift bin. 

Our analysis indicates that group-voids can reliably represent halo-voids in redshift bins with spectroscopic completeness of 40\% or higher, and that a completeness of at least 30\% provides an acceptable lower limit. 
We further demonstrated that adding a redshift error component to halos in simulations can successfully account for the impact of incompleteness, as well as the corresponding redshift uncertainties.
Importantly, using group-voids as a probe of the LSS remains fully suitable for spectroscopic surveys like DESI. For CSST, although obtaining highly precise redshifts for all of its more than 100 million galaxies is difficult, void statistics derived from group-void samples may provide an advantage over void analyses based directly on galaxy samples.

Finally, and most importantly, the direct link between group-void measurements and halo-voids in numerical simulations opens a new avenue for constructing a group (halo)-void emulator to facilitate cosmological research. Therefore, group-voids can serve as a potential and effective probe in LSS studies, with a superior advantage over traditional void analyses.

\begin{acknowledgments}

This work is supported by the National Key R\&D Program of China (2023YFA1607800, 2023YFA1607804, 2022YFA1605300), the China Manned Space Project with Nos. CMS-CSST-2021-A02 \& CMS-CSST-2025-A04, “the Fundamental Research Funds for the Central Universities”, 111 project No. B20019, and Shanghai Natural Science Foundation, grant No. 19ZR1466800, and the National Nature Science Foundation of China (NSFC) grants No. 12273051. This project is also supported in part by Office of Science and Technology, Shanghai Municipal Government (grant Nos. 24DX1400100, ZJ2023-ZD-001). Y.G. acknowledges the support from the CAS Project for Young Scientists in Basic Research (No. YSBR- 92), National Key R\&D Program of China grant Nos. 2022YFF0503404, and science research grants from the China Manned Space Project with grant Nos. CMS-CSST-2025-A02. Y.Z.G. acknowledges the support of the National Natural Science Foundation of China under the grant number 12503013. Y.C.Z. acknowledges the support of the National Natural Science Foundation of China through Grant Nos. 12273088. This work has made use of the Gravity Supercomputer at the Department of Astronomy, Shanghai Jiao Tong University.

\end{acknowledgments}



\bibliography{groupvoid}

@ARTICLE{gong,
       author = {{Gong}, Yan and {Liu}, Xiangkun and {Cao}, Ye and {Chen}, Xuelei and {Fan}, Zuhui and {Li}, Ran and {Li}, Xiao-Dong and {Li}, Zhigang and {Zhang}, Xin and {Zhan}, Hu},
        title = "{Cosmology from the Chinese Space Station Optical Survey (CSS-OS)}",
      journal = {\apj},
         year = 2019,
        month = oct,
       volume = {883},
       number = {2},
          eid = {203},
        pages = {203},
          doi = {10.3847/1538-4357/ab391e},
archivePrefix = {arXiv},
       eprint = {1901.04634},
 primaryClass = {astro-ph.CO},
       adsurl = {https://ui.adsabs.harvard.edu/abs/2019ApJ...883..203G}
}

@ARTICLE{zhan2021csst,
       author = {{Zhan}, Hu},
        title = "{The wide-field multiband imaging and slitless spectroscopy survey to be carried out by the Survey Space Telescope of China Manned Space Program}",
      journal = {Chinese Science Bulletin},
         year = 2021,
        month = apr,
       volume = {66},
       number = {11},
        pages = {1290},
          doi = {10.1360/TB-2021-0016},
}

@ARTICLE{2023MNRAS.519.1132M,
       author = {{Miao}, Haitao and {Gong}, Yan and {Chen}, Xuelei and {Huang}, Zhiqi and {Li}, Xiao-Dong and {Zhan}, Hu},
        title = "{Cosmological constraint precision of photometric and spectroscopic multi-probe surveys of China Space Station Telescope (CSST)}",
      journal = {\mnras},
         year = 2023,
        month = feb,
       volume = {519},
       number = {1},
        pages = {1132-1148},
          doi = {10.1093/mnras/stac3583},
archivePrefix = {arXiv},
       eprint = {2206.09822},
 primaryClass = {astro-ph.CO},
       adsurl = {https://ui.adsabs.harvard.edu/abs/2023MNRAS.519.1132M}
}

@ARTICLE{2025SCPMA..6880402G,
       author = {{Gong}, Yan and {Miao}, Haitao and {Zhou}, Xingchen and {Xiong}, Qi and {Song}, Yingxiao and {Jiang}, Yuer and {Wang}, Minglin and {Yan}, Junhui and {Wu}, Beichen and {Deng}, Furen and {Chen}, Xuelei and {Fan}, Zuhui and {Jing}, Yipeng and {Yang}, Xiaohu and {Zhan}, Hu},
        title = "{Future cosmology: New physics and opportunity from the China Space Station Telescope (CSST)}",
      journal = {Science China Physics, Mechanics, and Astronomy},
         year = 2025,
        month = aug,
       volume = {68},
       number = {8},
          eid = {280402},
        pages = {280402},
          doi = {10.1007/s11433-025-2646-2},
archivePrefix = {arXiv},
       eprint = {2501.15023},
 primaryClass = {astro-ph.CO},
       adsurl = {https://ui.adsabs.harvard.edu/abs/2025SCPMA..6880402G}
}

@ARTICLE{2026SCPMA..6939501C,
       author = {{CSST Collaboration} and {Gong}, Yan and {Miao}, Haitao and {Zhan}, Hu and {Li}, Zhao-Yu and {Shangguan}, Jinyi and {Li}, Haining and {Liu}, Chao and {Chen}, Xuefei and {Yuan}, Haibo and {Zhou}, Jilin and {Liu}, Hui-Gen and {Yu}, Cong and {Ji}, Jianghui and {Qi}, Zhaoxiang and {Liu}, Jiacheng and {Dai}, Zigao and {Wang}, Xiaofeng and {Zheng}, Zhenya and {Hao}, Lei and {Dou}, Jiangpei and {Ao}, Yiping and {Lin}, Zhenhui and {Zhang}, Kun and {Wang}, Wei and {Sun}, Guotong and {Li}, Ran and {Li}, Guoliang and {Xu}, Youhua and {Li}, Xinfeng and {Li}, Shengyang and {Wu}, Peng and {Zhang}, Jiuxing and {Wang}, Bo and {Bai}, Jinming and {Cai}, Yi-Fu and {Cai}, Zheng and {Cao}, Jie and {Chan}, Kwan Chuen and {Chang}, Jin and {Chen}, Xiaodian and {Chen}, Xuelei and {Chen}, Yuqin and {Chen}, Yun and {Cui}, Wei and {Dong}, Subo and {Du}, Pu and {Duan}, Wenying and {Fan}, Junhui and {Fan}, LuLu and {Fan}, Zhou and {Fan}, Zuhui and {Fang}, Taotao and {Fu}, Jianning and {Fu}, Liping and {Fu}, Zhensen and {Gao}, Jian and {Gu}, Shenghong and {Gu}, Yidong and {Guo}, Qi and {Han}, Zhanwen and {Hu}, Bin and {Huang}, Zhiqi and {Ho}, Luis C. and {Jiang}, Linhua and {Jiang}, Ning and {Jing}, Yipeng and {Kang}, Xi and {Kong}, Xu and {Li}, Cheng and {Li}, Chengyuan and {Li}, Di and {Li}, Jing and {Li}, Nan and {Li}, Yang A. and {Liao}, Shilong and {Lin}, Weipeng and {Liu}, Fengshan and {Liu}, Jifeng and {Liu}, Xiangkun and {Liu}, Zhuokai and {Mao}, Ruiqing and {Mao}, Shude and {Meng}, Xianmin and {Pang}, Xiaoying and {Peng}, Xiyan and {Peng}, Yingjie and {Shan}, Huanyuan and {Shen}, Juntai and {Shen}, Shiyin and {Shen}, Zhiqiang and {Shi}, Sheng-Cai and {Shi}, Yong and {Tan}, Siyuan and {Tian}, Hao and {Wang}, Jianmin and {Wang}, Jun-Xian and {Wang}, Xin and {Wang}, Yuting and {Wu}, Hong and {Wu}, Jingwen and {Wu}, Xuebing and {Xu}, Chun and {Xue}, Xiang-Xiang and {Xue}, Yongquan and {Yang}, Ji and {Yang}, Xiaohu and {Yao}, Qijun and {Yuan}, Fangting and {Yuan}, Zhen and {Zhang}, Jun and {Zhang}, Pengjie and {Zhang}, Tianmeng and {Zhang}, Wei and {Zhang}, Xin and {Zhao}, Gang and {Zhao}, Gongbo and {Zhong}, Hongen and {Zhong}, Jing and {Zhou}, Liyong and {Zhu}, Wei and {Zu}, Ying},
        title = "{Introduction to the Chinese Space Station Survey Telescope (CSST)}",
      journal = {Science China Physics, Mechanics, and Astronomy},
         year = 2026,
        month = jan,
       volume = {69},
       number = {3},
          eid = {239501},
        pages = {239501},
          doi = {10.1007/s11433-025-2809-0},
archivePrefix = {arXiv},
       eprint = {2507.04618},
 primaryClass = {astro-ph.IM},
       adsurl = {https://ui.adsabs.harvard.edu/abs/2026SCPMA..6939501C}
}

@ARTICLE{2020A&A...641A...6P,
       author = {{Planck Collaboration} and {Aghanim}, N. and {Akrami}, Y. and {Ashdown}, M. and {Aumont}, J. and {Baccigalupi}, C. and {Ballardini}, M. and {Banday}, A.~J. and {Barreiro}, R.~B. and {Bartolo}, N. and {Basak}, S. and {Battye}, R. and {Benabed}, K. and {Bernard}, J. -P. and {Bersanelli}, M. and {Bielewicz}, P. and {Bock}, J.~J. and {Bond}, J.~R. and {Borrill}, J. and {Bouchet}, F.~R. and {Boulanger}, F. and {Bucher}, M. and {Burigana}, C. and {Butler}, R.~C. and {Calabrese}, E. and {Cardoso}, J. -F. and {Carron}, J. and {Challinor}, A. and {Chiang}, H.~C. and {Chluba}, J. and {Colombo}, L.~P.~L. and {Combet}, C. and {Contreras}, D. and {Crill}, B.~P. and {Cuttaia}, F. and {de Bernardis}, P. and {de Zotti}, G. and {Delabrouille}, J. and {Delouis}, J. -M. and {Di Valentino}, E. and {Diego}, J.~M. and {Dor{\'e}}, O. and {Douspis}, M. and {Ducout}, A. and {Dupac}, X. and {Dusini}, S. and {Efstathiou}, G. and {Elsner}, F. and {En{\ss}lin}, T.~A. and {Eriksen}, H.~K. and {Fantaye}, Y. and {Farhang}, M. and {Fergusson}, J. and {Fernandez-Cobos}, R. and {Finelli}, F. and {Forastieri}, F. and {Frailis}, M. and {Fraisse}, A.~A. and {Franceschi}, E. and {Frolov}, A. and {Galeotta}, S. and {Galli}, S. and {Ganga}, K. and {G{\'e}nova-Santos}, R.~T. and {Gerbino}, M. and {Ghosh}, T. and {Gonz{\'a}lez-Nuevo}, J. and {G{\'o}rski}, K.~M. and {Gratton}, S. and {Gruppuso}, A. and {Gudmundsson}, J.~E. and {Hamann}, J. and {Handley}, W. and {Hansen}, F.~K. and {Herranz}, D. and {Hildebrandt}, S.~R. and {Hivon}, E. and {Huang}, Z. and {Jaffe}, A.~H. and {Jones}, W.~C. and {Karakci}, A. and {Keih{\"a}nen}, E. and {Keskitalo}, R. and {Kiiveri}, K. and {Kim}, J. and {Kisner}, T.~S. and {Knox}, L. and {Krachmalnicoff}, N. and {Kunz}, M. and {Kurki-Suonio}, H. and {Lagache}, G. and {Lamarre}, J. -M. and {Lasenby}, A. and {Lattanzi}, M. and {Lawrence}, C.~R. and {Le Jeune}, M. and {Lemos}, P. and {Lesgourgues}, J. and {Levrier}, F. and {Lewis}, A. and {Liguori}, M. and {Lilje}, P.~B. and {Lilley}, M. and {Lindholm}, V. and {L{\'o}pez-Caniego}, M. and {Lubin}, P.~M. and {Ma}, Y. -Z. and {Mac{\'\i}as-P{\'e}rez}, J.~F. and {Maggio}, G. and {Maino}, D. and {Mandolesi}, N. and {Mangilli}, A. and {Marcos-Caballero}, A. and {Maris}, M. and {Martin}, P.~G. and {Martinelli}, M. and {Mart{\'\i}nez-Gonz{\'a}lez}, E. and {Matarrese}, S. and {Mauri}, N. and {McEwen}, J.~D. and {Meinhold}, P.~R. and {Melchiorri}, A. and {Mennella}, A. and {Migliaccio}, M. and {Millea}, M. and {Mitra}, S. and {Miville-Desch{\^e}nes}, M. -A. and {Molinari}, D. and {Montier}, L. and {Morgante}, G. and {Moss}, A. and {Natoli}, P. and {N{\o}rgaard-Nielsen}, H.~U. and {Pagano}, L. and {Paoletti}, D. and {Partridge}, B. and {Patanchon}, G. and {Peiris}, H.~V. and {Perrotta}, F. and {Pettorino}, V. and {Piacentini}, F. and {Polastri}, L. and {Polenta}, G. and {Puget}, J. -L. and {Rachen}, J.~P. and {Reinecke}, M. and {Remazeilles}, M. and {Renzi}, A. and {Rocha}, G. and {Rosset}, C. and {Roudier}, G. and {Rubi{\~n}o-Mart{\'\i}n}, J.~A. and {Ruiz-Granados}, B. and {Salvati}, L. and {Sandri}, M. and {Savelainen}, M. and {Scott}, D. and {Shellard}, E.~P.~S. and {Sirignano}, C. and {Sirri}, G. and {Spencer}, L.~D. and {Sunyaev}, R. and {Suur-Uski}, A. -S. and {Tauber}, J.~A. and {Tavagnacco}, D. and {Tenti}, M. and {Toffolatti}, L. and {Tomasi}, M. and {Trombetti}, T. and {Valenziano}, L. and {Valiviita}, J. and {Van Tent}, B. and {Vibert}, L. and {Vielva}, P. and {Villa}, F. and {Vittorio}, N. and {Wandelt}, B.~D. and {Wehus}, I.~K. and {White}, M. and {White}, S.~D.~M. and {Zacchei}, A. and {Zonca}, A.},
        title = "{Planck 2018 results. VI. Cosmological parameters}",
      journal = {\aap},
         year = 2020,
        month = sep,
       volume = {641},
          eid = {A6},
        pages = {A6},
          doi = {10.1051/0004-6361/201833910},
archivePrefix = {arXiv},
       eprint = {1807.06209},
 primaryClass = {astro-ph.CO},
       adsurl = {https://ui.adsabs.harvard.edu/abs/2020A&A...641A...6P}
}

@ARTICLE{vast,
       author = {{Douglass}, Kelly and {Veyrat}, Dahlia and {O'Neill}, Stephen and {BenZvi}, Segev and {Zaidouni}, Fatima and {Guzzetti}, Michaela},
        title = "{VAST: the Void Analysis Software Toolkit}",
      journal = {The Journal of Open Source Software},
         year = 2022,
        month = sep,
       volume = {7},
       number = {77},
          eid = {4033},
        pages = {4033},
          doi = {10.21105/joss.04033},
       adsurl = {https://ui.adsabs.harvard.edu/abs/2022JOSS....7.4033D}
}

@ARTICLE{vide,
       author = {{Sutter}, P.~M. and {Lavaux}, G. and {Hamaus}, N. and {Pisani}, A. and {Wandelt}, B.~D. and {Warren}, M. and {Villaescusa-Navarro}, F. and {Zivick}, P. and {Mao}, Q. and {Thompson}, B.~B.},
        title = "{VIDE: The Void IDentification and Examination toolkit}",
      journal = {Astronomy and Computing},
         year = 2015,
        month = mar,
       volume = {9},
        pages = {1-9},
          doi = {10.1016/j.ascom.2014.10.002},
archivePrefix = {arXiv},
       eprint = {1406.1191},
 primaryClass = {astro-ph.CO},
       adsurl = {https://ui.adsabs.harvard.edu/abs/2015A&C.....9....1S}
}

@ARTICLE{zobov,
       author = {{Neyrinck}, Mark C.},
        title = "{ZOBOV: a parameter-free void-finding algorithm}",
      journal = {\mnras},
         year = 2008,
        month = jun,
       volume = {386},
       number = {4},
        pages = {2101-2109},
          doi = {10.1111/j.1365-2966.2008.13180.x},
archivePrefix = {arXiv},
       eprint = {0712.3049},
 primaryClass = {astro-ph},
       adsurl = {https://ui.adsabs.harvard.edu/abs/2008MNRAS.386.2101N}
}

@ARTICLE{watershed,
       author = {{Platen}, Erwin and {van de Weygaert}, Rien and {Jones}, Bernard J.~T.},
        title = "{A cosmic watershed: the WVF void detection technique}",
      journal = {\mnras},
         year = 2007,
        month = sep,
       volume = {380},
       number = {2},
        pages = {551-570},
          doi = {10.1111/j.1365-2966.2007.12125.x},
archivePrefix = {arXiv},
       eprint = {0706.2788},
 primaryClass = {astro-ph},
       adsurl = {https://ui.adsabs.harvard.edu/abs/2007MNRAS.380..551P}
}

@ARTICLE{2022ApJ...936..161W,
       author = {{Wang}, Jiaqi and {Yang}, Xiaohu and {Zhang}, Jun and {Li}, Hekun and {Fong}, Matthew and {Xu}, Haojie and {He}, Min and {Gu}, Yizhou and {Luo}, Wentao and {Dong}, Fuyu and {Wang}, Yirong and {Li}, Qingyang and {Katsianis}, Antonios and {Wang}, Haoran and {Shen}, Zhi and {Alonso Vaquero}, Pedro and {Liu}, Cong and {Huang}, Yiqi and {Liu}, Zhenjie},
        title = "{Halo Properties and Mass Functions of Groups/Clusters from the DESI Legacy Imaging Surveys DR9}",
      journal = {\apj},
         year = 2022,
        month = sep,
       volume = {936},
       number = {2},
          eid = {161},
        pages = {161},
          doi = {10.3847/1538-4357/ac8986},
archivePrefix = {arXiv},
       eprint = {2207.12771},
 primaryClass = {astro-ph.CO},
       adsurl = {https://ui.adsabs.harvard.edu/abs/2022ApJ...936..161W}
}

@ARTICLE{contarini2023cosmological,
       author = {{Contarini}, Sofia and {Pisani}, Alice and {Hamaus}, Nico and {Marulli}, Federico and {Moscardini}, Lauro and {Baldi}, Marco},
        title = "{Cosmological Constraints from the BOSS DR12 Void Size Function}",
      journal = {\apj},
         year = 2023,
        month = aug,
       volume = {953},
       number = {1},
          eid = {46},
        pages = {46},
          doi = {10.3847/1538-4357/acde54},
archivePrefix = {arXiv},
       eprint = {2212.03873},
 primaryClass = {astro-ph.CO},
       adsurl = {https://ui.adsabs.harvard.edu/abs/2023ApJ...953...46C}
}

@ARTICLE{2025MNRAS.539.1692Z,
       author = {{Zhang}, Youcai and {Yang}, Xiaohu and {Guo}, Hong and {Wang}, Peng and {Shi}, Feng},
        title = "{Galaxy and halo properties around cosmic filaments from Sloan Digital Sky Survey Data Release 7 and the ELUCID simulation}",
      journal = {\mnras},
         year = 2025,
        month = may,
       volume = {539},
       number = {2},
        pages = {1692-1705},
          doi = {10.1093/mnras/staf611},
archivePrefix = {arXiv},
       eprint = {2504.07367},
 primaryClass = {astro-ph.CO},
       adsurl = {https://ui.adsabs.harvard.edu/abs/2025MNRAS.539.1692Z}
}

@ARTICLE{2024ApJ...969...89T,
       author = {{Thiele}, Leander and {Massara}, Elena and {Pisani}, Alice and {Hahn}, ChangHoon and {Spergel}, David N. and {Ho}, Shirley and {Wandelt}, Benjamin},
        title = "{Neutrino Mass Constraint from an Implicit Likelihood Analysis of BOSS Voids}",
      journal = {\apj},
         year = 2024,
        month = jul,
       volume = {969},
       number = {2},
          eid = {89},
        pages = {89},
          doi = {10.3847/1538-4357/ad434e},
archivePrefix = {arXiv},
       eprint = {2307.07555},
 primaryClass = {astro-ph.CO},
       adsurl = {https://ui.adsabs.harvard.edu/abs/2024ApJ...969...89T}
}

@ARTICLE{2023MNRAS.522..152P,
       author = {{Pelliciari}, D. and {Contarini}, S. and {Marulli}, F. and {Moscardini}, L. and {Giocoli}, C. and {Lesci}, G.~F. and {Dolag}, K.},
        title = "{Exploring the cosmological synergy between galaxy cluster and cosmic void number counts}",
      journal = {\mnras},
         year = 2023,
        month = jun,
       volume = {522},
       number = {1},
        pages = {152-164},
          doi = {10.1093/mnras/stad956},
archivePrefix = {arXiv},
       eprint = {2210.07248},
 primaryClass = {astro-ph.CO},
       adsurl = {https://ui.adsabs.harvard.edu/abs/2023MNRAS.522..152P}
}

@ARTICLE{2024JCAP...10..079V,
       author = {{Verza}, Giovanni and {Carbone}, Carmelita and {Pisani}, Alice and {Porciani}, Cristiano and {Matarrese}, Sabino},
        title = "{The universal multiplicity function: counting haloes and voids}",
      journal = {\jcap},
         year = 2024,
        month = oct,
       volume = {2024},
       number = {10},
          eid = {079},
        pages = {079},
          doi = {10.1088/1475-7516/2024/10/079},
archivePrefix = {arXiv},
       eprint = {2401.14451},
 primaryClass = {astro-ph.CO},
       adsurl = {https://ui.adsabs.harvard.edu/abs/2024JCAP...10..079V}
}

@ARTICLE{2025ApJ...993..227V,
       author = {{Verza}, Giovanni and {Degni}, Giulia and {Pisani}, Alice and {Hamaus}, Nico and {Massara}, Elena and {Benson}, Andrew and {Escoffier}, St{\'e}phanie and {Wang}, Yun and {Zhai}, Zhongxu and {Dor{\'e}}, Olivier},
        title = "{Cosmology with Voids from the Nancy Grace Roman Space Telescope}",
      journal = {\apj},
         year = 2025,
        month = nov,
       volume = {993},
       number = {2},
          eid = {227},
        pages = {227},
          doi = {10.3847/1538-4357/ae07d9},
archivePrefix = {arXiv},
       eprint = {2410.19713},
 primaryClass = {astro-ph.CO},
       adsurl = {https://ui.adsabs.harvard.edu/abs/2025ApJ...993..227V}
}

@ARTICLE{2022MNRAS.516.4307W,
       author = {{Woodfinden}, Alex and {Nadathur}, Seshadri and {Percival}, Will J. and {Radinovic}, Sladana and {Massara}, Elena and {Winther}, Hans A.},
        title = "{Measurements of cosmic expansion and growth rate of structure from voids in the Sloan Digital Sky Survey between redshift 0.07 and 1.0}",
      journal = {\mnras},
         year = 2022,
        month = nov,
       volume = {516},
       number = {3},
        pages = {4307-4323},
          doi = {10.1093/mnras/stac2475},
archivePrefix = {arXiv},
       eprint = {2205.06258},
 primaryClass = {astro-ph.CO},
       adsurl = {https://ui.adsabs.harvard.edu/abs/2022MNRAS.516.4307W}
}

@ARTICLE{2023A&A...670A..47B,
       author = {{Bonici}, M. and {Carbone}, C. and {Davini}, S. and {Vielzeuf}, P. and {Paganin}, L. and {Cardone}, V. and {Hamaus}, N. and {Pisani}, A. and {Hawken}, A.~J. and {Kovacs}, A. and {Nadathur}, S. and {Contarini}, S. and {Verza}, G. and {Tutusaus}, I. and {Marulli}, F. and {Moscardini}, L. and {Aubert}, M. and {Giocoli}, C. and {Pourtsidou}, A. and {Camera}, S. and {Escoffier}, S. and {Caminata}, A. and {Di Domizio}, S. and {Martinelli}, M. and {Pallavicini}, M. and {Pettorino}, V. and {Sakr}, Z. and {Sapone}, D. and {Testera}, G. and {Tosi}, S. and {Yankelevich}, V. and {Amara}, A. and {Auricchio}, N. and {Baldi}, M. and {Bonino}, D. and {Branchini}, E. and {Brescia}, M. and {Brinchmann}, J. and {Capobianco}, V. and {Carretero}, J. and {Castellano}, M. and {Cavuoti}, S. and {Cledassou}, R. and {Congedo}, G. and {Conversi}, L. and {Copin}, Y. and {Corcione}, L. and {Courbin}, F. and {Cropper}, M. and {Da Silva}, A. and {Degaudenzi}, H. and {Douspis}, M. and {Dubath}, F. and {Duncan}, C.~A.~J. and {Dupac}, X. and {Dusini}, S. and {Ealet}, A. and {Farrens}, S. and {Ferriol}, S. and {Fosalba}, P. and {Frailis}, M. and {Franceschi}, E. and {Fumana}, M. and {G{\'o}mez-Alvarez}, P. and {Garilli}, B. and {Gillis}, B. and {Grazian}, A. and {Grupp}, F. and {Guzzo}, L. and {Haugan}, S.~V.~H. and {Holmes}, W. and {Hormuth}, F. and {Hornstrup}, A. and {Jahnke}, K. and {K{\"u}mmel}, M. and {Kermiche}, S. and {Kiessling}, A. and {Kilbinger}, M. and {Kunz}, M. and {Kurki-Suonio}, H. and {Laureijs}, R. and {Ligori}, S. and {Lilje}, P.~B. and {Lloro}, I. and {Maiorano}, E. and {Mansutti}, O. and {Marggraf}, O. and {Markovic}, K. and {Massey}, R. and {Medinaceli}, E. and {Melchior}, M. and {Meneghetti}, M. and {Meylan}, G. and {Moresco}, M. and {Munari}, E. and {Niemi}, S.~M. and {Padilla}, C. and {Paltani}, S. and {Pasian}, F. and {Pedersen}, K. and {Percival}, W.~J. and {Pires}, S. and {Polenta}, G. and {Poncet}, M. and {Popa}, L. and {Raison}, F. and {Rebolo}, R. and {Renzi}, A. and {Rhodes}, J. and {Rossetti}, E. and {Saglia}, R. and {Sartoris}, B. and {Scodeggio}, M. and {Secroun}, A. and {Seidel}, G. and {Sirignano}, C. and {Sirri}, G. and {Stanco}, L. and {Starck}, J. -L. and {Surace}, C. and {Tallada-Cresp{\'\i}}, P. and {Tavagnacco}, D. and {Taylor}, A.~N. and {Tereno}, I. and {Toledo-Moreo}, R. and {Torradeflot}, F. and {Valentijn}, E.~A. and {Valenziano}, L. and {Wang}, Y. and {Weller}, J. and {Zamorani}, G. and {Zoubian}, J. and {Andreon}, S.},
        title = "{Euclid: Forecasts from the void-lensing cross-correlation}",
      journal = {\aap},
         year = 2023,
        month = feb,
       volume = {670},
          eid = {A47},
        pages = {A47},
          doi = {10.1051/0004-6361/202244445},
archivePrefix = {arXiv},
       eprint = {2206.14211},
 primaryClass = {astro-ph.CO},
       adsurl = {https://ui.adsabs.harvard.edu/abs/2023A&A...670A..47B}
}

@ARTICLE{2023JCAP...08..010V,
       author = {{Vielzeuf}, Pauline and {Calabrese}, Matteo and {Carbone}, Carmelita and {Fabbian}, Giulio and {Baccigalupi}, Carlo},
        title = "{DEMNUni: the imprint of massive neutrinos on the cross-correlation between cosmic voids and CMB lensing}",
      journal = {\jcap},
         year = 2023,
        month = aug,
       volume = {2023},
       number = {8},
          eid = {010},
        pages = {010},
          doi = {10.1088/1475-7516/2023/08/010},
archivePrefix = {arXiv},
       eprint = {2303.10048},
 primaryClass = {astro-ph.CO},
       adsurl = {https://ui.adsabs.harvard.edu/abs/2023JCAP...08..010V}
}

@ARTICLE{2023A&A...674A.185M,
       author = {{Mauland}, R. and {Elgar{\o}y}, {\O}. and {Mota}, D.~F. and {Winther}, H.~A.},
        title = "{The void-galaxy cross-correlation function with massive neutrinos and modified gravity}",
      journal = {\aap},
         year = 2023,
        month = jun,
       volume = {674},
          eid = {A185},
        pages = {A185},
          doi = {10.1051/0004-6361/202346287},
archivePrefix = {arXiv},
       eprint = {2303.05820},
 primaryClass = {astro-ph.CO},
       adsurl = {https://ui.adsabs.harvard.edu/abs/2023A&A...674A.185M}
}

@ARTICLE{2023A&A...677A..78R,
       author = {{Radinovi{\'c}}, S. and {Nadathur}, S. and {Winther}, H. -A. and {Percival}, W.~J. and {Woodfinden}, A. and {Massara}, E. and {Paillas}, E. and {Contarini}, S. and {Hamaus}, N. and {Kovacs}, A. and {Pisani}, A. and {Verza}, G. and {Aubert}, M. and {Amara}, A. and {Auricchio}, N. and {Baldi}, M. and {Bonino}, D. and {Branchini}, E. and {Brescia}, M. and {Camera}, S. and {Capobianco}, V. and {Carbone}, C. and {Cardone}, V.~F. and {Carretero}, J. and {Castellano}, M. and {Cavuoti}, S. and {Cimatti}, A. and {Cledassou}, R. and {Congedo}, G. and {Conversi}, L. and {Copin}, Y. and {Corcione}, L. and {Courbin}, F. and {Da Silva}, A. and {Douspis}, M. and {Dubath}, F. and {Dupac}, X. and {Farrens}, S. and {Ferriol}, S. and {Fosalba}, P. and {Frailis}, M. and {Franceschi}, E. and {Fumana}, M. and {Galeotta}, S. and {Garilli}, B. and {Gillard}, W. and {Gillis}, B. and {Giocoli}, C. and {Grazian}, A. and {Grupp}, F. and {Haugan}, S.~V.~H. and {Holmes}, W. and {Hornstrup}, A. and {Jahnke}, K. and {K{\"u}mmel}, M. and {Kiessling}, A. and {Kilbinger}, M. and {Kitching}, T. and {Kurki-Suonio}, H. and {Ligori}, S. and {Lilje}, P.~B. and {Lloro}, I. and {Maiorano}, E. and {Mansutti}, O. and {Marggraf}, O. and {Markovic}, K. and {Marulli}, F. and {Massey}, R. and {Mei}, S. and {Melchior}, M. and {Mellier}, Y. and {Meneghetti}, M. and {Merlin}, E. and {Meylan}, G. and {Moresco}, M. and {Moscardini}, L. and {Niemi}, S. -M. and {Nightingale}, J.~W. and {Nutma}, T. and {Padilla}, C. and {Paltani}, S. and {Pasian}, F. and {Pedersen}, K. and {Pettorino}, V. and {Pires}, S. and {Polenta}, G. and {Poncet}, M. and {Popa}, L.~A. and {Pozzetti}, L. and {Raison}, F. and {Renzi}, A. and {Rhodes}, J. and {Riccio}, G. and {Romelli}, E. and {Roncarelli}, M. and {Rosset}, C. and {Saglia}, R. and {Sapone}, D. and {Sartoris}, B. and {Schneider}, P. and {Secroun}, A. and {Seidel}, G. and {Serrano}, S. and {Sirignano}, C. and {Sirri}, G. and {Stanco}, L. and {Starck}, J. -L. and {Surace}, C. and {Tallada-Cresp{\'\i}}, P. and {Tereno}, I. and {Toledo-Moreo}, R. and {Torradeflot}, F. and {Tutusaus}, I. and {Valentijn}, E.~A. and {Valenziano}, L. and {Vassallo}, T. and {Wang}, Y. and {Weller}, J. and {Zamorani}, G. and {Zoubian}, J. and {Scottez}, V.},
        title = "{Euclid: Cosmology forecasts from the void-galaxy cross-correlation function with reconstruction}",
      journal = {\aap},
         year = 2023,
        month = sep,
       volume = {677},
          eid = {A78},
        pages = {A78},
          doi = {10.1051/0004-6361/202346121},
archivePrefix = {arXiv},
       eprint = {2302.05302},
 primaryClass = {astro-ph.CO},
       adsurl = {https://ui.adsabs.harvard.edu/abs/2023A&A...677A..78R}
}

@ARTICLE{2022A&A...667A.162C,
       author = {{Contarini}, S. and {Verza}, G. and {Pisani}, A. and {Hamaus}, N. and {Sahl{\'e}n}, M. and {Carbone}, C. and {Dusini}, S. and {Marulli}, F. and {Moscardini}, L. and {Renzi}, A. and {Sirignano}, C. and {Stanco}, L. and {Aubert}, M. and {Bonici}, M. and {Castignani}, G. and {Courtois}, H.~M. and {Escoffier}, S. and {Guinet}, D. and {Kovacs}, A. and {Lavaux}, G. and {Massara}, E. and {Nadathur}, S. and {Pollina}, G. and {Ronconi}, T. and {Ruppin}, F. and {Sakr}, Z. and {Veropalumbo}, A. and {Wandelt}, B.~D. and {Amara}, A. and {Auricchio}, N. and {Baldi}, M. and {Bonino}, D. and {Branchini}, E. and {Brescia}, M. and {Brinchmann}, J. and {Camera}, S. and {Capobianco}, V. and {Carretero}, J. and {Castellano}, M. and {Cavuoti}, S. and {Cledassou}, R. and {Congedo}, G. and {Conselice}, C.~J. and {Conversi}, L. and {Copin}, Y. and {Corcione}, L. and {Courbin}, F. and {Cropper}, M. and {Da Silva}, A. and {Degaudenzi}, H. and {Dubath}, F. and {Duncan}, C.~A.~J. and {Dupac}, X. and {Ealet}, A. and {Farrens}, S. and {Ferriol}, S. and {Fosalba}, P. and {Frailis}, M. and {Franceschi}, E. and {Garilli}, B. and {Gillard}, W. and {Gillis}, B. and {Giocoli}, C. and {Grazian}, A. and {Grupp}, F. and {Guzzo}, L. and {Haugan}, S. and {Holmes}, W. and {Hormuth}, F. and {Jahnke}, K. and {K{\"u}mmel}, M. and {Kermiche}, S. and {Kiessling}, A. and {Kilbinger}, M. and {Kunz}, M. and {Kurki-Suonio}, H. and {Laureijs}, R. and {Ligori}, S. and {Lilje}, P.~B. and {Lloro}, I. and {Maiorano}, E. and {Mansutti}, O. and {Marggraf}, O. and {Markovic}, K. and {Massey}, R. and {Melchior}, M. and {Meneghetti}, M. and {Meylan}, G. and {Moresco}, M. and {Munari}, E. and {Niemi}, S.~M. and {Padilla}, C. and {Paltani}, S. and {Pasian}, F. and {Pedersen}, K. and {Percival}, W.~J. and {Pettorino}, V. and {Pires}, S. and {Polenta}, G. and {Poncet}, M. and {Popa}, L. and {Pozzetti}, L. and {Raison}, F. and {Rhodes}, J. and {Rossetti}, E. and {Saglia}, R. and {Sartoris}, B. and {Schneider}, P. and {Secroun}, A. and {Seidel}, G. and {Sirri}, G. and {Surace}, C. and {Tallada-Cresp{\'\i}}, P. and {Taylor}, A.~N. and {Tereno}, I. and {Toledo-Moreo}, R. and {Torradeflot}, F. and {Valentijn}, E.~A. and {Valenziano}, L. and {Wang}, Y. and {Weller}, J. and {Zamorani}, G. and {Zoubian}, J. and {Andreon}, S. and {Maino}, D. and {Mei}, S.},
        title = "{Euclid: Cosmological forecasts from the void size function}",
      journal = {\aap},
         year = 2022,
        month = nov,
       volume = {667},
          eid = {A162},
        pages = {A162},
          doi = {10.1051/0004-6361/202244095},
archivePrefix = {arXiv},
       eprint = {2205.11525},
 primaryClass = {astro-ph.CO},
       adsurl = {https://ui.adsabs.harvard.edu/abs/2022A&A...667A.162C}
}

@ARTICLE{2016ApJ...833..241S,
       author = {{Shi}, Feng and {Yang}, Xiaohu and {Wang}, Huiyuan and {Zhang}, Youcai and {Mo}, H.~J. and {van den Bosch}, Frank C. and {Li}, Shijie and {Liu}, Chengze and {Lu}, Yi and {Tweed}, Dylan and {Yang}, Lei},
        title = "{Mapping the Real-space Distributions of Galaxies in SDSS DR7. I. Two-point Correlation Functions}",
      journal = {\apj},
         year = 2016,
        month = dec,
       volume = {833},
       number = {2},
          eid = {241},
        pages = {241},
          doi = {10.3847/1538-4357/833/2/241},
archivePrefix = {arXiv},
       eprint = {1608.02313},
 primaryClass = {astro-ph.CO},
       adsurl = {https://ui.adsabs.harvard.edu/abs/2016ApJ...833..241S}
}

@ARTICLE{2018ApJ...861..137S,
       author = {{Shi}, Feng and {Yang}, Xiaohu and {Wang}, Huiyuan and {Zhang}, Youcai and {Mo}, H.~J. and {van den Bosch}, Frank C. and {Luo}, Wentao and {Tweed}, Dylan and {Li}, Shijie and {Liu}, Chengze and {Lu}, Yi and {Yang}, Lei},
        title = "{Mapping the Real Space Distributions of Galaxies in SDSS DR7. II. Measuring the Growth Rate, Clustering Amplitude of Matter, and Biases of Galaxies at Redshift 0.1}",
      journal = {\apj},
         year = 2018,
        month = jul,
       volume = {861},
       number = {2},
          eid = {137},
        pages = {137},
          doi = {10.3847/1538-4357/aacb20},
archivePrefix = {arXiv},
       eprint = {1712.04163},
 primaryClass = {astro-ph.CO},
       adsurl = {https://ui.adsabs.harvard.edu/abs/2018ApJ...861..137S}
}

@ARTICLE{2025MNRAS.540.2853S,
       author = {{Song}, Yingxiao and {Gong}, Yan and {Zhou}, Xingchen and {Miao}, Haitao and {Chan}, Kwan Chuen and {Chen}, Xuelei},
        title = "{Cosmological constraints using the void size function data from BOSS DR16}",
      journal = {\mnras},
         year = 2025,
        month = jul,
       volume = {540},
       number = {3},
        pages = {2853-2862},
          doi = {10.1093/mnras/staf918},
archivePrefix = {arXiv},
       eprint = {2501.07817},
 primaryClass = {astro-ph.CO},
       adsurl = {https://ui.adsabs.harvard.edu/abs/2025MNRAS.540.2853S}
}

@ARTICLE{2024MNRAS.532.1049S,
       author = {{Song}, Yingxiao and {Xiong}, Qi and {Gong}, Yan and {Deng}, Furen and {Chan}, Kwan Chuen and {Chen}, Xuelei and {Guo}, Qi and {Han}, Jiaxin and {Li}, Guoliang and {Li}, Ming and {Liu}, Yun and {Luo}, Yu and {Pei}, Wenxiang and {Wei}, Chengliang},
        title = "{Cosmological forecast of the void size function measurement from the CSST spectroscopic survey}",
      journal = {\mnras},
         year = 2024,
        month = jul,
       volume = {532},
       number = {1},
        pages = {1049-1058},
          doi = {10.1093/mnras/stae1575},
archivePrefix = {arXiv},
       eprint = {2402.05492},
 primaryClass = {astro-ph.CO},
       adsurl = {https://ui.adsabs.harvard.edu/abs/2024MNRAS.532.1049S}
}

@ARTICLE{2024MNRAS.534..128S,
       author = {{Song}, Yingxiao and {Xiong}, Qi and {Gong}, Yan and {Deng}, Furen and {Chan}, Kwan Chuen and {Chen}, Xuelei and {Guo}, Qi and {Liu}, Yun and {Pei}, Wenxiang},
        title = "{Void number counts as a cosmological probe for the large-scale structure}",
      journal = {\mnras},
         year = 2024,
        month = oct,
       volume = {534},
       number = {1},
        pages = {128-134},
          doi = {10.1093/mnras/stae2094},
archivePrefix = {arXiv},
       eprint = {2409.03178},
 primaryClass = {astro-ph.CO},
       adsurl = {https://ui.adsabs.harvard.edu/abs/2024MNRAS.534..128S}
}

@ARTICLE{contarini2021cosmic,
       author = {{Contarini}, Sofia and {Marulli}, Federico and {Moscardini}, Lauro and {Veropalumbo}, Alfonso and {Giocoli}, Carlo and {Baldi}, Marco},
        title = "{Cosmic voids in modified gravity models with massive neutrinos}",
      journal = {\mnras},
         year = 2021,
        month = jul,
       volume = {504},
       number = {4},
        pages = {5021-5038},
          doi = {10.1093/mnras/stab1112},
archivePrefix = {arXiv},
       eprint = {2009.03309},
 primaryClass = {astro-ph.CO},
       adsurl = {https://ui.adsabs.harvard.edu/abs/2021MNRAS.504.5021C}
}

@ARTICLE{2017MNRAS.469..787P,
       author = {{Pollina}, Giorgia and {Hamaus}, Nico and {Dolag}, Klaus and {Weller}, Jochen and {Baldi}, Marco and {Moscardini}, Lauro},
        title = "{On the linearity of tracer bias around voids}",
      journal = {\mnras},
         year = 2017,
        month = jul,
       volume = {469},
       number = {1},
        pages = {787-799},
          doi = {10.1093/mnras/stx785},
archivePrefix = {arXiv},
       eprint = {1610.06176},
 primaryClass = {astro-ph.CO},
       adsurl = {https://ui.adsabs.harvard.edu/abs/2017MNRAS.469..787P}
}

@ARTICLE{2019MNRAS.487.2836P,
       author = {{Pollina}, G. and {Hamaus}, N. and {Paech}, K. and {Dolag}, K. and {Weller}, J. and {S{\'a}nchez}, C. and {Rykoff}, E.~S. and {Jain}, B. and {Abbott}, T.~M.~C. and {Allam}, S. and {Avila}, S. and {Bernstein}, R.~A. and {Bertin}, E. and {Brooks}, D. and {Burke}, D.~L. and {Carnero Rosell}, A. and {Carrasco Kind}, M. and {Carretero}, J. and {Cunha}, C.~E. and {D'Andrea}, C.~B. and {da Costa}, L.~N. and {De Vicente}, J. and {DePoy}, D.~L. and {Desai}, S. and {Diehl}, H.~T. and {Doel}, P. and {Evrard}, A.~E. and {Flaugher}, B. and {Fosalba}, P. and {Frieman}, J. and {Garc{\'\i}a-Bellido}, J. and {Gerdes}, D.~W. and {Giannantonio}, T. and {Gruen}, D. and {Gschwend}, J. and {Gutierrez}, G. and {Hartley}, W.~G. and {Hollowood}, D.~L. and {Honscheid}, K. and {Hoyle}, B. and {James}, D.~J. and {Jeltema}, T. and {Kuehn}, K. and {Kuropatkin}, N. and {Lima}, M. and {March}, M. and {Marshall}, J.~L. and {Melchior}, P. and {Menanteau}, F. and {Miquel}, R. and {Plazas}, A.~A. and {Romer}, A.~K. and {Sanchez}, E. and {Scarpine}, V. and {Schindler}, R. and {Schubnell}, M. and {Sevilla-Noarbe}, I. and {Smith}, M. and {Soares-Santos}, M. and {Sobreira}, F. and {Suchyta}, E. and {Tarle}, G. and {Walker}, A.~R. and {Wester}, W. and {DES Collaboration}},
        title = "{On the relative bias of void tracers in the Dark Energy Survey}",
      journal = {\mnras},
         year = 2019,
        month = aug,
       volume = {487},
       number = {2},
        pages = {2836-2852},
          doi = {10.1093/mnras/stz1470},
archivePrefix = {arXiv},
       eprint = {1806.06860},
 primaryClass = {astro-ph.CO},
       adsurl = {https://ui.adsabs.harvard.edu/abs/2019MNRAS.487.2836P}
}

@ARTICLE{2025MNRAS.538..114S,
       author = {{Song}, Yingxiao and {Gong}, Yan and {Xiong}, Qi and {Chan}, Kwan Chuen and {Chen}, Xuelei and {Guo}, Qi and {Liu}, Yun and {Pei}, Wenxiang},
        title = "{2D watershed void clustering for probing the cosmic large-scale structure}",
      journal = {\mnras},
         year = 2025,
        month = mar,
       volume = {538},
       number = {1},
        pages = {114-120},
          doi = {10.1093/mnras/staf305},
archivePrefix = {arXiv},
       eprint = {2410.04898},
 primaryClass = {astro-ph.CO},
       adsurl = {https://ui.adsabs.harvard.edu/abs/2025MNRAS.538..114S}
}

@ARTICLE{2024ApJ...976..244S,
       author = {{Song}, Yingxiao and {Xiong}, Qi and {Gong}, Yan and {Deng}, Furen and {Chan}, Kwan Chuen and {Chen}, Xuelei and {Guo}, Qi and {Li}, Guoliang and {Li}, Ming and {Liu}, Yun and {Luo}, Yu and {Pei}, Wenxiang and {Wei}, Chengliang},
        title = "{Cosmological Prediction of the Void and Galaxy Clustering Measurements in the CSST Spectroscopic Survey}",
      journal = {\apj},
         year = 2024,
        month = dec,
       volume = {976},
       number = {2},
          eid = {244},
        pages = {244},
          doi = {10.3847/1538-4357/ad8de9},
archivePrefix = {arXiv},
       eprint = {2408.08589},
 primaryClass = {astro-ph.CO},
       adsurl = {https://ui.adsabs.harvard.edu/abs/2024ApJ...976..244S}
}

@ARTICLE{2022A&A...658A..20H,
       author = {{Hamaus}, N. and {Aubert}, M. and {Pisani}, A. and {Contarini}, S. and {Verza}, G. and {Cousinou}, M. -C. and {Escoffier}, S. and {Hawken}, A. and {Lavaux}, G. and {Pollina}, G. and {Wandelt}, B.~D. and {Weller}, J. and {Bonici}, M. and {Carbone}, C. and {Guzzo}, L. and {Kovacs}, A. and {Marulli}, F. and {Massara}, E. and {Moscardini}, L. and {Ntelis}, P. and {Percival}, W.~J. and {Radinovi{\'c}}, S. and {Sahl{\'e}n}, M. and {Sakr}, Z. and {S{\'a}nchez}, A.~G. and {Winther}, H.~A. and {Auricchio}, N. and {Awan}, S. and {Bender}, R. and {Bodendorf}, C. and {Bonino}, D. and {Branchini}, E. and {Brescia}, M. and {Brinchmann}, J. and {Capobianco}, V. and {Carretero}, J. and {Castander}, F.~J. and {Castellano}, M. and {Cavuoti}, S. and {Cimatti}, A. and {Cledassou}, R. and {Congedo}, G. and {Conversi}, L. and {Copin}, Y. and {Corcione}, L. and {Cropper}, M. and {Da Silva}, A. and {Degaudenzi}, H. and {Douspis}, M. and {Dubath}, F. and {Duncan}, C.~A.~J. and {Dupac}, X. and {Dusini}, S. and {Ealet}, A. and {Ferriol}, S. and {Fosalba}, P. and {Frailis}, M. and {Franceschi}, E. and {Franzetti}, P. and {Fumana}, M. and {Garilli}, B. and {Gillis}, B. and {Giocoli}, C. and {Grazian}, A. and {Grupp}, F. and {Haugan}, S.~V.~H. and {Holmes}, W. and {Hormuth}, F. and {Jahnke}, K. and {Kermiche}, S. and {Kiessling}, A. and {Kilbinger}, M. and {Kitching}, T. and {K{\"u}mmel}, M. and {Kunz}, M. and {Kurki-Suonio}, H. and {Ligori}, S. and {Lilje}, P.~B. and {Lloro}, I. and {Maiorano}, E. and {Marggraf}, O. and {Markovic}, K. and {Massey}, R. and {Maurogordato}, S. and {Melchior}, M. and {Meneghetti}, M. and {Meylan}, G. and {Moresco}, M. and {Munari}, E. and {Niemi}, S.~M. and {Padilla}, C. and {Paltani}, S. and {Pasian}, F. and {Pedersen}, K. and {Pettorino}, V. and {Pires}, S. and {Poncet}, M. and {Popa}, L. and {Pozzetti}, L. and {Rebolo}, R. and {Rhodes}, J. and {Rix}, H. and {Roncarelli}, M. and {Rossetti}, E. and {Saglia}, R. and {Schneider}, P. and {Secroun}, A. and {Seidel}, G. and {Serrano}, S. and {Sirignano}, C. and {Sirri}, G. and {Starck}, J. -L. and {Tallada-Cresp{\'\i}}, P. and {Tavagnacco}, D. and {Taylor}, A.~N. and {Tereno}, I. and {Toledo-Moreo}, R. and {Torradeflot}, F. and {Valentijn}, E.~A. and {Valenziano}, L. and {Wang}, Y. and {Welikala}, N. and {Zamorani}, G. and {Zoubian}, J. and {Andreon}, S. and {Baldi}, M. and {Camera}, S. and {Mei}, S. and {Neissner}, C. and {Romelli}, E.},
        title = "{Euclid: Forecasts from redshift-space distortions and the Alcock-Paczynski test with cosmic voids}",
      journal = {\aap},
         year = 2022,
        month = feb,
       volume = {658},
          eid = {A20},
        pages = {A20},
          doi = {10.1051/0004-6361/202142073},
archivePrefix = {arXiv},
       eprint = {2108.10347},
 primaryClass = {astro-ph.CO},
       adsurl = {https://ui.adsabs.harvard.edu/abs/2022A&A...658A..20H}
}

@ARTICLE{2024A&A...691A..39R,
       author = {{Radinovi{\'c}}, Sla{\dj}ana and {Winther}, Hans A. and {Nadathur}, Seshadri and {Percival}, Will J. and {Paillas}, Enrique and {Sohrab Fraser}, Tristan and {Massara}, Elena and {Woodfinden}, Alex},
        title = "{Alcock{\textendash}Paczy{\'n}ski effect on void-finding: Implications for void-galaxy cross-correlation modelling}",
      journal = {\aap},
         year = 2024,
        month = nov,
       volume = {691},
          eid = {A39},
        pages = {A39},
          doi = {10.1051/0004-6361/202451358},
archivePrefix = {arXiv},
       eprint = {2407.02699},
 primaryClass = {astro-ph.CO},
       adsurl = {https://ui.adsabs.harvard.edu/abs/2024A&A...691A..39R}
}

@ARTICLE{2022MNRAS.513..186A,
       author = {{Aubert}, Marie and {Cousinou}, Marie-Claude and {Escoffier}, St{\'e}phanie and {Hawken}, Adam J. and {Nadathur}, Seshadri and {Alam}, Shadab and {Bautista}, Julian and {Burtin}, Etienne and {Chuang}, Chia-Hsun and {de la Macorra}, Axel and {de Mattia}, Arnaud and {Gil-Mar{\'\i}n}, H{\'e}ctor and {Hou}, Jiamin and {Jullo}, Eric and {Kneib}, Jean-Paul and {Neveux}, Richard and {Rossi}, Graziano and {Schneider}, Donald and {Smith}, Alex and {Tamone}, Am{\'e}lie and {Vargas Maga{\~n}a}, Mariana and {Zhao}, Cheng},
        title = "{The completed SDSS-IV extended Baryon Oscillation Spectroscopic Survey: growth rate of structure measurement from cosmic voids}",
      journal = {\mnras},
         year = 2022,
        month = jun,
       volume = {513},
       number = {1},
        pages = {186-203},
          doi = {10.1093/mnras/stac828},
archivePrefix = {arXiv},
       eprint = {2007.09013},
 primaryClass = {astro-ph.CO},
       adsurl = {https://ui.adsabs.harvard.edu/abs/2022MNRAS.513..186A}
}

@ARTICLE{2016MNRAS.462.2465C,
       author = {{Cai}, Yan-Chuan and {Taylor}, Andy and {Peacock}, John A. and {Padilla}, Nelson},
        title = "{Redshift-space distortions around voids}",
      journal = {\mnras},
         year = 2016,
        month = nov,
       volume = {462},
       number = {3},
        pages = {2465-2477},
          doi = {10.1093/mnras/stw1809},
archivePrefix = {arXiv},
       eprint = {1603.05184},
 primaryClass = {astro-ph.CO},
       adsurl = {https://ui.adsabs.harvard.edu/abs/2016MNRAS.462.2465C}
}

@ARTICLE{2019MNRAS.483.3472N,
       author = {{Nadathur}, Seshadri and {Percival}, Will J.},
        title = "{An accurate linear model for redshift space distortions in the void-galaxy correlation function}",
      journal = {\mnras},
         year = 2019,
        month = mar,
       volume = {483},
       number = {3},
        pages = {3472-3487},
          doi = {10.1093/mnras/sty3372},
archivePrefix = {arXiv},
       eprint = {1712.07575},
 primaryClass = {astro-ph.CO},
       adsurl = {https://ui.adsabs.harvard.edu/abs/2019MNRAS.483.3472N}
}

@ARTICLE{pisani2015counting,
       author = {{Pisani}, Alice and {Sutter}, P.~M. and {Hamaus}, Nico and {Alizadeh}, Esfandiar and {Biswas}, Rahul and {Wandelt}, Benjamin D. and {Hirata}, Christopher M.},
        title = "{Counting voids to probe dark energy}",
      journal = {\prd},
         year = 2015,
        month = oct,
       volume = {92},
       number = {8},
          eid = {083531},
        pages = {083531},
          doi = {10.1103/PhysRevD.92.083531},
archivePrefix = {arXiv},
       eprint = {1503.07690},
 primaryClass = {astro-ph.CO},
       adsurl = {https://ui.adsabs.harvard.edu/abs/2015PhRvD..92h3531P}
}

@ARTICLE{2017MNRAS.465..746S,
       author = {{S{\'a}nchez}, C. and {Clampitt}, J. and {Kovacs}, A. and {Jain}, B. and {Garc{\'\i}a-Bellido}, J. and {Nadathur}, S. and {Gruen}, D. and {Hamaus}, N. and {Huterer}, D. and {Vielzeuf}, P. and {Amara}, A. and {Bonnett}, C. and {DeRose}, J. and {Hartley}, W.~G. and {Jarvis}, M. and {Lahav}, O. and {Miquel}, R. and {Rozo}, E. and {Rykoff}, E.~S. and {Sheldon}, E. and {Wechsler}, R.~H. and {Zuntz}, J. and {Abbott}, T.~M.~C. and {Abdalla}, F.~B. and {Annis}, J. and {Benoit-L{\'e}vy}, A. and {Bernstein}, G.~M. and {Bernstein}, R.~A. and {Bertin}, E. and {Brooks}, D. and {Buckley-Geer}, E. and {Carnero Rosell}, A. and {Carrasco Kind}, M. and {Carretero}, J. and {Crocce}, M. and {Cunha}, C.~E. and {D'Andrea}, C.~B. and {da Costa}, L.~N. and {Desai}, S. and {Diehl}, H.~T. and {Dietrich}, J.~P. and {Doel}, P. and {Evrard}, A.~E. and {Fausti Neto}, A. and {Flaugher}, B. and {Fosalba}, P. and {Frieman}, J. and {Gaztanaga}, E. and {Gruendl}, R.~A. and {Gutierrez}, G. and {Honscheid}, K. and {James}, D.~J. and {Krause}, E. and {Kuehn}, K. and {Lima}, M. and {Maia}, M.~A.~G. and {Marshall}, J.~L. and {Melchior}, P. and {Plazas}, A.~A. and {Reil}, K. and {Romer}, A.~K. and {Sanchez}, E. and {Schubnell}, M. and {Sevilla-Noarbe}, I. and {Smith}, R.~C. and {Soares-Santos}, M. and {Sobreira}, F. and {Suchyta}, E. and {Tarle}, G. and {Thomas}, D. and {Walker}, A.~R. and {Weller}, J. and {DES Collaboration}},
        title = "{Cosmic voids and void lensing in the Dark Energy Survey Science Verification data}",
      journal = {\mnras},
         year = 2017,
        month = feb,
       volume = {465},
       number = {1},
        pages = {746-759},
          doi = {10.1093/mnras/stw2745},
archivePrefix = {arXiv},
       eprint = {1605.03982},
 primaryClass = {astro-ph.CO},
       adsurl = {https://ui.adsabs.harvard.edu/abs/2017MNRAS.465..746S}
}

@ARTICLE{2021MNRAS.500..464V,
       author = {{Vielzeuf}, P. and {Kov{\'a}cs}, A. and {Demirbozan}, U. and {Fosalba}, P. and {Baxter}, E. and {Hamaus}, N. and {Huterer}, D. and {Miquel}, R. and {Nadathur}, S. and {Pollina}, G. and {S{\'a}nchez}, C. and {Whiteway}, L. and {Abbott}, T.~M.~C. and {Allam}, S. and {Annis}, J. and {Avila}, S. and {Brooks}, D. and {Burke}, D.~L. and {Carnero Rosell}, A. and {Carrasco Kind}, M. and {Carretero}, J. and {Cawthon}, R. and {Costanzi}, M. and {da Costa}, L.~N. and {De Vicente}, J. and {Desai}, S. and {Diehl}, H.~T. and {Doel}, P. and {Eifler}, T.~F. and {Everett}, S. and {Flaugher}, B. and {Frieman}, J. and {Garc{\'\i}a-Bellido}, J. and {Gaztanaga}, E. and {Gerdes}, D.~W. and {Gruen}, D. and {Gruendl}, R.~A. and {Gschwend}, J. and {Gutierrez}, G. and {Hartley}, W.~G. and {Hollowood}, D.~L. and {Honscheid}, K. and {James}, D.~J. and {Kuehn}, K. and {Kuropatkin}, N. and {Lahav}, O. and {Lima}, M. and {Maia}, M.~A.~G. and {March}, M. and {Marshall}, J.~L. and {Melchior}, P. and {Menanteau}, F. and {Palmese}, A. and {Paz-Chinch{\'o}n}, F. and {Plazas}, A.~A. and {Sanchez}, E. and {Scarpine}, V. and {Serrano}, S. and {Sevilla-Noarbe}, I. and {Smith}, M. and {Suchyta}, E. and {Tarle}, G. and {Thomas}, D. and {Weller}, J. and {Zuntz}, J. and {Zuntz}, J. and {DES Collaboration}},
        title = "{Dark Energy Survey Year 1 results: the lensing imprint of cosmic voids on the cosmic microwave background}",
      journal = {\mnras},
         year = 2021,
        month = jan,
       volume = {500},
       number = {1},
        pages = {464-480},
          doi = {10.1093/mnras/staa3231},
archivePrefix = {arXiv},
       eprint = {1911.02951},
 primaryClass = {astro-ph.CO},
       adsurl = {https://ui.adsabs.harvard.edu/abs/2021MNRAS.500..464V}
}

@ARTICLE{2018MNRAS.476.3195C,
       author = {{Cautun}, Marius and {Paillas}, Enrique and {Cai}, Yan-Chuan and {Bose}, Sownak and {Armijo}, Joaquin and {Li}, Baojiu and {Padilla}, Nelson},
        title = "{The Santiago-Harvard-Edinburgh-Durham void comparison - I. SHEDding light on chameleon gravity tests}",
      journal = {\mnras},
         year = 2018,
        month = may,
       volume = {476},
       number = {3},
        pages = {3195-3217},
          doi = {10.1093/mnras/sty463},
archivePrefix = {arXiv},
       eprint = {1710.01730},
 primaryClass = {astro-ph.CO},
       adsurl = {https://ui.adsabs.harvard.edu/abs/2018MNRAS.476.3195C}
}

@ARTICLE{2005MNRAS.356.1293Y,
       author = {{Yang}, Xiaohu and {Mo}, H.~J. and {van den Bosch}, Frank C. and {Jing}, Y.~P.},
        title = "{A halo-based galaxy group finder: calibration and application to the 2dFGRS}",
      journal = {\mnras},
         year = 2005,
        month = feb,
       volume = {356},
       number = {4},
        pages = {1293-1307},
          doi = {10.1111/j.1365-2966.2005.08560.x},
archivePrefix = {arXiv},
       eprint = {astro-ph/0405234},
 primaryClass = {astro-ph},
       adsurl = {https://ui.adsabs.harvard.edu/abs/2005MNRAS.356.1293Y}
}

@ARTICLE{2007ApJ...671..153Y,
       author = {{Yang}, Xiaohu and {Mo}, H.~J. and {van den Bosch}, Frank C. and {Pasquali}, Anna and {Li}, Cheng and {Barden}, Marco},
        title = "{Galaxy Groups in the SDSS DR4. I. The Catalog and Basic Properties}",
      journal = {\apj},
         year = 2007,
        month = dec,
       volume = {671},
       number = {1},
        pages = {153-170},
          doi = {10.1086/522027},
archivePrefix = {arXiv},
       eprint = {0707.4640},
 primaryClass = {astro-ph},
       adsurl = {https://ui.adsabs.harvard.edu/abs/2007ApJ...671..153Y}
}

@ARTICLE{2012ApJ...752...41Y,
       author = {{Yang}, Xiaohu and {Mo}, H.~J. and {van den Bosch}, Frank C. and {Zhang}, Youcai and {Han}, Jiaxin},
        title = "{Evolution of the Galaxy-Dark Matter Connection and the Assembly of Galaxies in Dark Matter Halos}",
      journal = {\apj},
         year = 2012,
        month = jun,
       volume = {752},
       number = {1},
          eid = {41},
        pages = {41},
          doi = {10.1088/0004-637X/752/1/41},
archivePrefix = {arXiv},
       eprint = {1110.1420},
 primaryClass = {astro-ph.CO},
       adsurl = {https://ui.adsabs.harvard.edu/abs/2012ApJ...752...41Y}
}

@ARTICLE{2021ApJ...909..143Y,
       author = {{Yang}, Xiaohu and {Xu}, Haojie and {He}, Min and {Gu}, Yizhou and {Katsianis}, Antonios and {Meng}, Jiacheng and {Shi}, Feng and {Zou}, Hu and {Zhang}, Youcai and {Liu}, Chengze and {Wang}, Zhaoyu and {Dong}, Fuyu and {Lu}, Yi and {Li}, Qingyang and {Chen}, Yangyao and {Wang}, Huiyuan and {Mo}, Houjun and {Fu}, Jian and {Guo}, Hong and {Leauthaud}, Alexie and {Luo}, Yu and {Zhang}, Jun and {Zu}, Ying},
        title = "{An Extended Halo-based Group/Cluster Finder: Application to the DESI Legacy Imaging Surveys DR8}",
      journal = {\apj},
         year = 2021,
        month = mar,
       volume = {909},
       number = {2},
          eid = {143},
        pages = {143},
          doi = {10.3847/1538-4357/abddb2},
archivePrefix = {arXiv},
       eprint = {2012.14998},
 primaryClass = {astro-ph.GA},
       adsurl = {https://ui.adsabs.harvard.edu/abs/2021ApJ...909..143Y}
}
\bibliographystyle{aasjournal}

\end{document}